\documentclass{ian}

\usepackage{graphicx}
\usepackage{array}
\usepackage{largearray}
\usepackage{dsfont}
\usepackage{iantheo}
\usepackage[reset_at_theo]{point}

\begin{document}

\pagestyle{empty}

\hbox{}
\hfil{\bf\LARGE
The Hierarchical Graphene model\par
}
\vfill

\hfil{\bf\large Ian Jauslin}\par
\hfil{\it Department of Mathematics, Rutgers University}\par
\hfil{\tt\color{blue}\href{mailto:ian.jauslin@rutgers.edu}{ian.jauslin@rutgers.edu}}\par
\vskip20pt

\vfill

{\it\small
  These are the lecture notes for the summer school {\rm``Quantum Mechanics from Condensed Matter to Computing''}, organized by {\rm Niels Benedikter, Marcin Napi\'orkowski, Jan Philip Solovej} and {\rm Albert Werner} in Copenhagen from June 13 to 17, 2022.
}

\vfill

\hfil {\bf Abstract}\par
{\small
The hierarchical graphene model is a simple toy model which is useful to understand the mechanics of renormalization group flows in super-renormalizable systems.
It is based on a model of interacting electrons in graphene, for which the renormalization group analysis was carried out by Giuliani and Mastropietro.
The analysis of the hierarchical graphene model is significantly simpler than graphene, but one should not expect it to produce good quantitative results about real-world graphene.
Rather, the hierarchical model is useful as a teaching tool to understand the core concepts of renormalization group techniques.
In this paper, we will first introduce a model for electrons in graphene and set it up for a renormalization group treatment by introducing its Grassmann representation and scale decomposition.
We then define the hierarchical graphene model and study it's renormalization group flow.
From a renormalization group point of view, graphene is quite simple: it is super-renormalizable.
As an illustration of a more complicated system, we repeat the analysis for the Kondo model, which is a strongly coupled model with a non-trivial fixed point.
}

\vfill

\tableofcontents

\vfill
\eject

\setcounter{page}1
\pagestyle{plain}

\section{Introduction}
\indent
The renormalization group is a powerful technique used to study a wide variety of systems: from field theories to statistical mechanical systems.
The technique comes in many different flavors, some more appropriate to numerical computations, others tailored to theoretical analyses.
Some can be justified mathematically, whereas for others, understanding why they work is still an open problem.
In this paper, our focus will be purposefully narrow, and we will focus on the {\it Wilsonian} renormalization group\-~\cite{Wi65}, which, to put it simply, consists in separately considering different energy scales, and studying how systems behave in each one.
We will only discuss Fermionic statistical field theories, that is, systems of many Fermions, for which mathematically complete analyses abound.
Our point of view will be that developed by Benfatto and Gallavotti\-~\cite{BG90}, which has been applied successfully to a variety of systems: to name but a few, \cite{BGe94,Ma11,GMP12,GGM12,GMP17} among many others.
\bigskip

\indent
Of particular interest in this paper will be a model for interacting electrons in graphene.
Graphene is a two-dimensional crystal of carbon atoms in a honeycomb structure, whose discovery in 2007 set off a flurry of interest\-~\cite{NGe04} due to its unusual and potentially useful properties.
To start out with, two dimensional crystals (without a substrate) are quite rare.
This two-dimensional structure makes graphene crystals atom-thin and very flexible, but since the carbon atoms in graphene are bound covalently, it can sustain very high stresses.
It is also an extremely good conductor.
These properties give graphene great potential in many technological applications, from flexible displays to lightweight, conducting and tear-resistant plastics.
\bigskip

\indent
Graphene is often studied in the approximation that its electrons do not interact with each other\-~\cite{Mc57,SW58}.
In this case, the electronic properties of graphene can be computed exactly.
However, taking into account interactions between electrons is more complicated.
We will use the renormalization group to accomplish just this, and show that the interactions, if they are weak enough, do not change the physical picture much.
In other words, the renormalization group allows us to set up a perturbation theory for the electrons in graphene, where the perturbation is the interaction.
Perturbation theory in many-Fermion systems is relatively straightforward when the unperturbed Hamiltonian is ``gapped''.
This is not the case with graphene, so standard perturbation theory does not work.
However, the gap closes only at two points in momentum space, called Fermi points, and near these Fermi points, the bands are approximately conical.
Both of these properties make graphene, from a renormalization group point of view, {\it super-renormalizable}, which one can understand as ``easier to study using the renormalization group than generic models''.
This is the prime motivation for studying this model in this paper: it will serve as an example in which to understand the core concepts of Wilsonian renormalization.
\bigskip

\indent
The renormalization group analysis of interacting graphene was carried out in\-~\cite{GM10,Gi10}, and was used to prove the universality of the conductivity\-~\cite{GMP12}.
As is apparent from the length of these papers, despite the relative simplicity of graphene, it is still not a trivial task to study it using the renormalization group.
In order to keep the discussion as simple as possible and nevertheless get to the core ideas of the renormalization group, we will simplify the graphene model, and introduce the {\it hierarchical graphene model}.
Hierarchical models have long been used:\-~\cite{Dy69,BCe78,GK81} as toy models in which to understand renormalization group analyses in a simpler setting.
The renormalization group is notorious for requiring a large amount of mathematical details to be worked out.
Hierarchical models can be studied without so many difficulties, and can help to grasp the conceptual core of the renormalization group.
Hierarchical models for Fermionic systems, first studied by Dorlas\-~\cite{Do91}, are even simpler than Bosonic ones: as was shown in\-~\cite{BGJ15,GJ15}, they are {\it integrable}, in that their renormalization group analysis can be carried out {\it explicitly} (in Bosonic and non-hierarchical cases, the renormalization group produces infinite power series).
It is important to emphasize that hierarchical models are {\it toy models}: they are not approximations of their non-hierarchical counterparts, nor do they make good quantitative predictions about them.
They are only really useful to understand renormalization group flows on a conceptual level.
\bigskip

\indent
For this paper, we will define the hierarchical graphene model, following the ideas of\-~\cite{BGJ15}.
(This model has not been introduced before, though this is more likely because it is more useful as a teaching tool than as a physical model.)
The main idea is to eliminate everything from the graphene model except for its scaling properties.
This will leave us with a simple model, for which we can compute the renormalization group flow {\it exactly} and {\it explicitly}, which will help us understand what a renormlization group flow is, and how it can be used.
We will briefly discuss how to adapt the analysis to non-hierarchical graphene, but few details will be given.
Interested readers are referred to the detailed presentation in\-~\cite{Gi10}.
\bigskip

\indent
The hierarchical graphene model will allow us to get a handle on the renormalization group analysis for graphene, which, as was mentioned above, is a {\it perturbative} analysis (by which we mean that the interaction is a perturbation; the term ``perturbative'' is sometimes used to mean ``formal'', but this is not what is meant here).
Conceptually, the renormalization group has much to say about non-perturbative systems.
For instance, consider a system of electrons in a superconducting phase.
On short length scales (high energies), the electrons are essentially independent from each other, but on large length scales (low energies), they form {\it Cooper pairs}, which allows the system to conduct electricity without resistance.
From a renormalization group point of view, we should see that at small distances, interactions between electrons are not important, but at large distances, they change the behavior of electrons {\it qualitatively}.
In other words, the effective model at large length scales should be very different from the non-interacting one (in fact, one should see BCS theory emerging at these scales).
However, the renormalization group is so difficult to study away from the perturbative regime that this has, so far, never been accomplished (at least not mathematically): not for BCS theory, nor for {\it any} other strongly coupled system.
\bigskip

\indent
However, Fermionic hierarchical models can be studied exactly using the renormalization group, even in strongly coupled situations.
We give an example of such a system at the end of this paper: the hierarchical Kondo model.
The Kondo model is a one-dimensional system of electrons on a lattice that interact with a localizaed magnetic impurity.
It was introduced\-~\cite{Ko64} as a toy model to study conductance in disordered systems.
It was studied rather extensively by Anderson\-~\cite{An70}, and later found to be exactly solvable by Andrei\-~\cite{An80}.
In developing his version of the renormalization group, Wilson studied the Kondo model\-~\cite{Wi75}, though his analysis has not yet been made mathematically rigorous.
One way to see that the Kondo model is strongly coupled is through the {\it Kondo effect}, in which the magnetic impurity can be shown to have a finite susceptibility at zero temperature.
This means that, in the lowest energy state, if one applies a magnetic field to the impurity, it will not align perfectly with the field.
In the absence of interactions with the electrons, this would obviously not be true (the susceptibility would be infinite).
But, even if the interaction is arbitrarily small, as long as it is ferromagnetic, the susceptibility comes out finite.
\bigskip

\indent
The Kondo effect can be seen in the renormalization group flow, which, at small energy scales, goes to a {\it non-trivial fixed point}, that is, to an effective theory that is qualitatively different from the non-interacting one.
This is similar, in essence, to the BCS question described above.
Whereas the Kondo effect can be proved by using the exact solvability of the Kondo model\-~\cite{An80}, it has never been shown using the renormalization group.
It is, however, provable in the hierarchical Kondo model\-~\cite{BGJ15,GJ15}, which will be briefly discussed at the end of this paper.
\bigskip

\indent
The rest of this paper is structured as follows.
In section\-~\ref{sec:graphene}, we introduce the (non-hierarchical) graphene model and set up its renormalization group analysis by expressing observables in terms of Grassmann variables and decomposing the system into scales.
In section\-~\ref{sec:hierarchical_graphene}, we define the hierarchical graphene model and study its renormalization group flow.
In section\-~\ref{sec:hierarchical_kondo}, we define the hierarchical Kondo model and discuss its renormalization group flow.
There are two appendices in which useful lemmas are proved: in appendix\-~\ref{app:free_fermions} we compute properties of free Fermion systems, and in appendix\-~\ref{app:grassmann}, we prove some properties of Gaussian Grassmann integrals.

\section{Graphene}\label{sec:graphene}
\indent
In this section, we describe a model for the electrons in graphene.
We will use the {\it tight-binding} approximation, in which the electrons are assumed to be bound to carbon atoms on a hexagonal lattice (see figure\-~\ref{fig:graphene}).
For more details on this model, see\-~\cite{Gi10,GM10}.
\bigskip

\begin{figure}
\hfil\includegraphics[width=12cm]{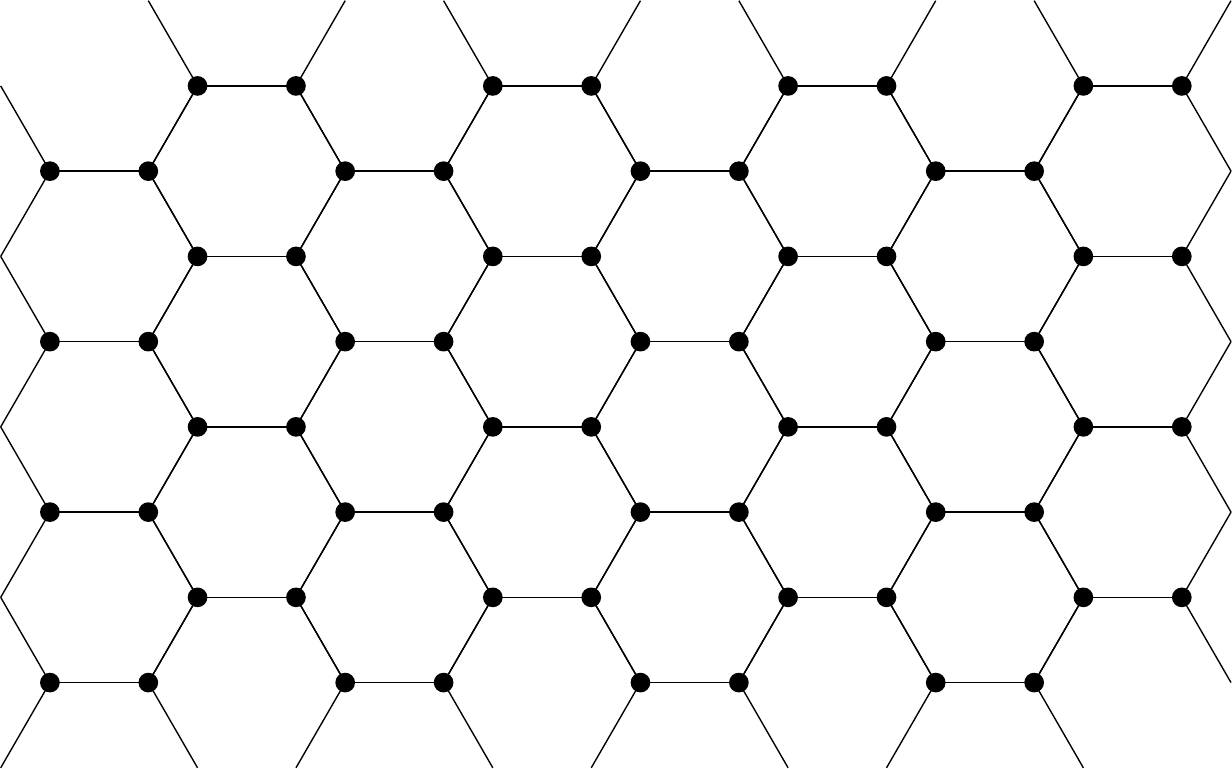}
\caption{
  The hexagonal lattice.
  Each dot represents a carbon atom, and the lines connect neighboring atoms, and representing the possible hoppings paths for the electrons.
}
\label{fig:graphene}
\end{figure}

\subsection{Lattice}
\indent
The hexagonal lattice can be constructed by copying an elementary cell at every integer combination of
\begin{equation}
  l_1:=\left(\frac{3}{2},\frac{\sqrt{3}}{2}\right)
  ,\quad
  l_2:=\left(\frac{3}{2},-\frac{\sqrt{3}}{2}\right)
  \label{laeo}
\end{equation}
where we have chosen the unit length to be equal to the distance between two nearest neighbors.
The elementary cell consists of two atoms at $(0,0)$ and at $(0,1)$ (relative to the position of the cell).
\bigskip

\begin{figure}
\hfil\includegraphics[width=12cm]{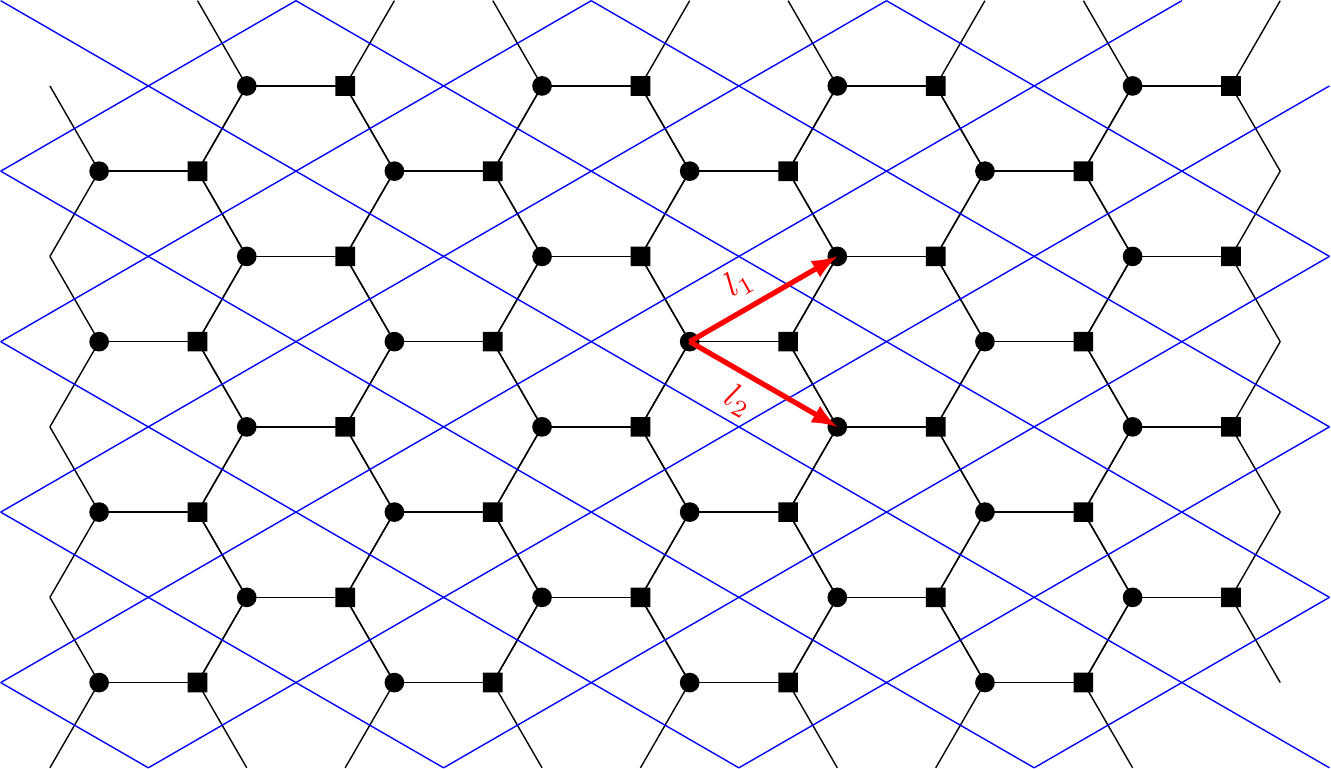}
\caption{
  The cell decomposition of the hexagonal lattice.
  Each cell (blue rhombus) contains two atoms.
  One of type $a$ (circle) and one of type $b$ (square).
}
\label{fig:graphene}
\end{figure}

\indent
We define the lattice 
\begin{equation}
  \Lambda:=\left\{n_1 l_1+n_2 l_2,\ (n_1,n_2)\in\{0,\cdots,L-1\}^2\right\}
  \label{laet}
\end{equation}
where $L$ is a positive integer that determines the size of the crystal, that we will eventually send to infinity, with periodic boundary conditions.
We introduce the nearest neighbor vectors:
\begin{equation}
  \delta_1:=(1,0)
  ,\quad
  \delta_2:=\left(-\frac{1}{2},\frac{\sqrt{3}}{2}\right)
  ,\quad
  \delta_3:=\left(-\frac{1}{2},-\frac{\sqrt{3}}{2}\right).
  \label{lea}
\end{equation}
\bigskip

\begin{figure}
\hfil\includegraphics[width=4cm]{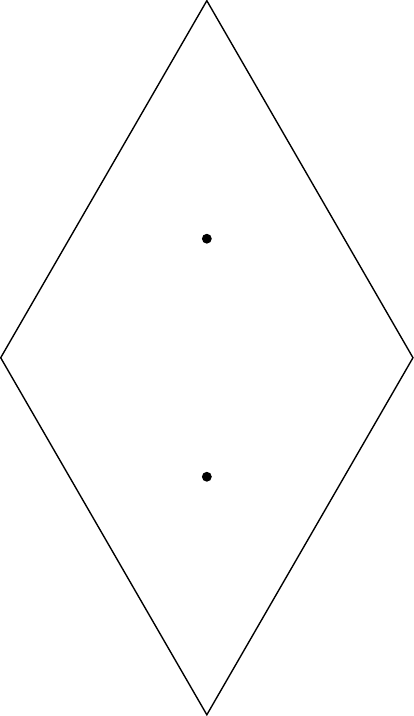}
\caption{
  The Brillouin zone.
  The two dots are the Fermi points $p_F^{(\pm)}$, see below.
}
\label{fig:brillouin}
\end{figure}

\indent
The {\it dual} of $\Lambda$ is
\begin{equation}
  \hat\Lambda:=\left\{\frac{m_1}{L} G_1+\frac{m_2}{L} G_2,\ (m_1,m_2)\in\{0,\cdots,L-1\}^2\right\}
  \label{lae}
\end{equation}
with periodic boundary conditions, where
\begin{equation}
  G_1=\left(\frac{2\pi}{3},\frac{2\pi}{\sqrt{3}}\right)
  ,\quad
  G_2=\left(\frac{2\pi}{3},-\frac{2\pi}{\sqrt{3}}\right).
  \label{laeG}
\end{equation}
It is defined in such a way that $\forall x\in\Lambda$, $\forall k\in\hat\Lambda$, 
\begin{equation}
  e^{ikxL}=1.
\end{equation}
In the limit $L\to\infty$, the set $\hat \Lambda$ tends to the torus $\hat \Lambda_\infty=\mathbb R^2/(\mathbb Z G_1+\mathbb Z G_2)$, also called the {\it Brillouin zone}, see figure\-~\ref{fig:brillouin}.
\bigskip

\subsection{Hamiltonian}
\indent
We will consider a model of spin-$\frac12$ electrons on the hexagonal lattice.
Given $ x\in\Lambda$, we denote the Fermionic annihilation operators with spin $\sigma$ at atoms of type $a$ and $b$ within the elementary cell centered at $x$ respectively by $a_{x,\sigma}$ and $b_{x+\delta_1,\sigma}$.
\bigskip

\indent
The Hamiltonian is split into two terms:
\begin{equation}
  \mathcal H=\mathcal H_0+\mathcal H_I
\end{equation}
where $\mathcal H_0$ is the {\it free Hamiltonian}, which describes the motion of electrons from one atom to a neighbor, and $\mathcal H_I$ is the {\it interaction Hamiltonian}, which describes the interaction between electrons.
\bigskip

\begin{figure}
\hfil\includegraphics[width=8cm, trim={1cm 1cm 1cm 1cm}, clip]{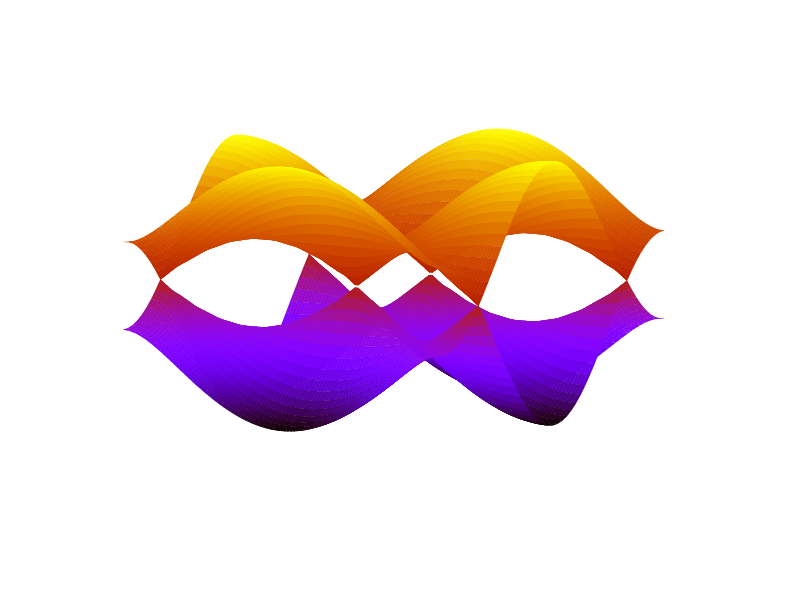}
\caption{
  The bands of graphene.
  The two bands meet at two points in a conical intersection.
}
\label{fig:bands}
\end{figure}

\point{\bf Free Hamiltonian.}
The free Hamiltonian describes the {\it hopping} of electrons from one atom to another:
\begin{equation}
  \mathcal H_0:=-\sum_{\sigma\in\{\uparrow,\downarrow\}}\sum_{\displaystyle\mathop{\scriptstyle x\in\Lambda}_{j=1,2,3}}(a_{x,\sigma}^\dagger b_{x +\delta_j,\sigma}+b_{x +\delta_j,\sigma}^\dagger a_{x,\sigma})
  \label{hamx}
\end{equation}
Equation\-~(\ref{hamx}) can be rewritten in Fourier space as follows.
We define the Fourier transform of the annihilation operators as
\begin{equation}
  \hat a_{k,\sigma}:=\frac1{\sqrt{|\Lambda|}}\sum_{x\in\Lambda}e^{ikx}a_{x,\sigma}
  ,\quad 
  \hat b_{k,\sigma}:=\frac1{\sqrt{|\Lambda|}}\sum_{x\in\Lambda}e^{ikx}b_{x+\delta_1,\sigma}
\end{equation} 
where $|\Lambda|=L^2$.
Note that, with this choice of normalization, $\hat a_{k,\sigma}$ and $\hat b_{k,\sigma}$ satisfy the canonical anticommutation relations:
\begin{equation}
  \{a_{k,\sigma},a_{k',\sigma'}^\dagger\}
  =
  \frac1{|\Lambda|}\sum_{x,x'\in\Lambda}e^{ikx-ik'x'}
  \{a_{x,\sigma},a_{x',\sigma'}^\dagger\}
  =
  \delta_{\sigma,\sigma'}
  \frac1{|\Lambda|}\sum_{x'\in\Lambda}e^{i(k-k')x}
  =
  \delta_{\sigma,\sigma'}
  \delta_{k,k'}
\end{equation}
and similarly for $b$.
We express $\mathcal H_0$ in terms of $\hat a$ and $\hat b$:
\begin{equation}
  \mathcal H_0=-\sum_{\sigma\in\{\uparrow,\downarrow\}}\sum_{ k\in\hat\Lambda}\hat A_{k,\sigma}^\dagger H_0(k)\hat A_{k,\sigma}
  \label{hamk}
\end{equation}
$\hat A_{k,\sigma}$ is a column vector whose transpose is $\hat A_{k,\sigma}^T=(\hat a_{k,\sigma},\hat{b}_{k,\sigma})$, 
\begin{equation}
  H_0( k):=
  \left(\begin{array}{*{2}{c}}
    0&\Omega^*(k)\\
    \Omega(k)&0
  \end{array}\right)
  \label{hmat}
\end{equation}
and
\begin{equation}
  \Omega(k):=\sum_{j=1}^3e^{ik(\delta_j-\delta_1)}
  =1+2e^{-i\frac32k_x}\cos({\textstyle\frac{\sqrt{3}}2k_y})
  .
  \label{Omega}
\end{equation}
The eigenvalues of $H_0(k)$ are called the {\it bands} of non-interacting graphene, and are
\begin{equation}
  \pm|\Omega(k)|
  =\pm\left(
    1+4\cos({\textstyle\frac32k_x})\cos({\textstyle\frac{\sqrt{3}}2k_y})
    +4\cos^2({\textstyle\frac{\sqrt{3}}2k_y})
  \right)^{\frac12}
  .
  \label{bands}
\end{equation}
These bands meet at $0$ at exactly two values of $k$: for $\omega=\pm$,
\begin{equation}
  p_F^{(\omega)}:=({\textstyle\frac{2\pi}3,\omega\frac{2\pi}{3\sqrt3}})
  \label{fermipt}
\end{equation}
see figure\-~\ref{fig:bands}.
For $|k-p_F^{(\omega)}|\ll1$,
\begin{equation}
  \pm|\Omega(k)|\sim\pm v_F|k-p_F^{(\omega)}|
  ,\quad
  v_F=\frac32
  .
  \label{dispersion}
\end{equation}
\bigskip

\bigskip

\point{\bf Interaction.}
We now define the interaction Hamiltonian which we take to be of {\it Hubbard} form:
\begin{equation}
  \mathcal H_I:=
  U\sum_{x\in\Lambda}\sum_{\alpha\in\{a,b\}}\left(\alpha_{x+d_\alpha,\uparrow}^\dagger\alpha_{x+d_\alpha,\uparrow}-\frac{1}{2}\right)\left(\alpha_{x+d_\alpha,\downarrow}^\dagger\alpha_{x+d_\alpha,\downarrow}-\frac{1}{2}\right)
\label{hamintx}\end{equation}
where the $d_\alpha$ are the vectors that give the position of each atom type with respect to the centers of the lattice $\Lambda$: $d_a:=0$, $d_b:=\delta_1$.

\subsection{Grassmann integral representation}
\indent
The {\it specific free energy} on the lattice $\Lambda$ is defined by
\begin{equation}
  f_\Lambda:=-\frac{1}{\beta|\Lambda|}\log\left(\mathrm{Tr}\left( e^{-\beta\mathcal H}\right)\right)
  \label{freeen}
\end{equation}
where $\beta$ is the inverse temperature.
We define these at finite $\beta$ and $L$, but will take $\beta,L\to\infty$.
A straightforward application of the Trotter product formula implies that (see\-~\cite[(4.1)]{Gi10})
\begin{equation}
  \mathrm{Tr}( e^{-\beta\mathcal H})
  =
  \mathrm{Tr}(e^{-\beta\mathcal H_0})
  +\sum_{N=1}^\infty\frac{(-\beta)^N}{N!}\int_{\beta\geqslant t_1\geqslant\cdots\geqslant t_N\geqslant0}\mathrm{Tr}\left(e^{-\beta\mathcal H_0}\mathcal H_I(t_1)\cdots\mathcal H_I(t_N)\right)
\end{equation}
where
\begin{equation}
  \mathcal H_I(t):=e^{t\mathcal H_0}\mathcal H_Ie^{-t\mathcal H_0}
  .
\end{equation}
To compute this trace, we will use the {\it Wick rule}, which we will now descibe.
First, we define the {\it free average}:
\begin{equation}
  \left<A\right>:=\frac{\mathrm{Tr}(e^{-\beta\mathcal H_0}A)}{\mathrm{Tr}(e^{-\beta\mathcal H_0})}
  .
\end{equation}
Next, we define the imaginary time creation and annihilation operators: for $\alpha\in\{a,b\}$ and $t\in[0,\beta)$,
\begin{equation}
  \alpha_{x,\sigma}^{+}(t):=e^{t\mathcal H_0}\alpha_{x,\sigma}^\dagger e^{-t\mathcal H_0}
  ,\quad
  \alpha_{x,\sigma}^{-}(t):=e^{t\mathcal H_0}\alpha_{x,\sigma} e^{-t\mathcal H_0}
\end{equation}
(note that $\alpha^+$ is not the adjoint of $\alpha^-$).
The Wick rule can be used to compute the free average of any polynomial of the creation and annihilation operators: it is linear, and, for any $n\in\mathbb N$, $\alpha^{(1)},\bar\alpha^{(1)},\cdots,\alpha^{(n)},\bar\alpha^{(n)}\in\{a,b\}$ $x_1,\bar x_1,\cdots,x_{n},\bar x_n\in\Lambda$, $\sigma_1,\bar\sigma_1,\cdots,\sigma_n,\bar\sigma_n\in\{\uparrow,\downarrow\}$, $\beta\geqslant t_1\geqslant \bar t_1>\cdots>t_{n}\geqslant \bar t_n\geqslant0$,
\begin{equation}
  \left<\prod_{i=1}^{n}\alpha^{(i)-}_{x_i,\sigma_i}(t_i)\bar\alpha^{(i)+}_{\bar x_{i},\bar\sigma_i}(\bar t_{i})\right>
  =
  \sum_{\tau\in\mathcal S_n}(-1)^\tau\prod_{i=1}^{n}\left<\mathbf T\left(\alpha^{(i)-}_{x_i,\sigma_i}(t_i)\bar\alpha^{(\tau(i))+}_{\bar x_{\tau(i)},\bar\sigma_{\tau(i)}}(\bar t_{\tau(i)})\right)\right>
  \label{wick}
\end{equation}
(to alleviate the notation, we have replaced $\alpha_{x+d_\alpha}$ by $\alpha_x$ here) where $\mathcal S_n$ is the set of permutations of $\{1,\cdots,n\}$, $(-1)^\tau$ is the signature of $\tau$ and $\mathbf T$ is the time-ordering operator:
\begin{equation}
  \mathbf T\left(\alpha^-_{x,\sigma}(t)\bar\alpha^+_{\bar x,\bar\sigma}(\bar t)\right)
  =\left\{\begin{array}{>\displaystyle ll}
    \alpha^-_{x,\sigma}(t)\bar\alpha^+_{\bar x,\bar\sigma}(\bar t)
    &\mathrm{if\ }t\geqslant \bar t
    \\
    -\bar\alpha^+_{\bar x,\bar\sigma}(\bar t)\alpha^-_{x,\sigma}(t)
    &\mathrm{if\ }t< \bar t
    .
  \end{array}\right.
\end{equation}
The Wick rule can be proved by a direct computation, and follows from the fact that $\mathcal H_0$ is quadratic in the annihilation operators, see lemma\-~\ref{lemma:wick}.
\bigskip

\indent
Fermionic creation and annihilation operators do not anticommute: $\{a_i,a_i\dagger\}=1$.
However, the time-ordering operator effectively makes them anticommute.
We can make this more precise be re-expressing the problem in terms of {\it Grassmann variables}.
\bigskip

\point{\bf Definition of the Grassmann algebra.}
We first define a Grassmann algebra and an integration procedure on it.
We will work in Fourier space, in both space and time.
Let us first define the Fourier transform in time: for every $\alpha\in\{a,b\}$, $\sigma\in\{\uparrow,\downarrow\}$ for $k_0\in2\pi\beta^{-1}(\mathbb Z+1/2)$, $k\in\hat\Lambda$, and $\mathbf k\equiv(k_0,k)$
\begin{equation}
  \hat\alpha_{\mathbf k,\sigma}^\pm:=\int_0^\beta dt\ e^{\mp itk_0}e^{\mathcal H_0 t}\hat\alpha_{k,\sigma}^\pm e^{-\mathcal H_0 t}
  \equiv
  \frac1{\sqrt{|\Lambda|}}\sum_{x\in\Lambda}\int_0^\beta dt\ e^{\mp(itk_0+ikx)}\alpha_{x,\sigma}^\pm(t)
\end{equation}
in which we use the shorthand $\hat\alpha_{k,\sigma}^-\equiv\hat\alpha_{k,\sigma}$, $\hat\alpha_{k,\sigma}^+\equiv\hat\alpha_{k,\sigma}^\dagger$.
We notice that $\mathbf k\in\mathcal B_{\beta,L}:=(2\pi\beta^{-1}(\mathbb Z+1/2))\times\hat\Lambda$ varies in an infinite set.
Since this will cause trouble when defining Grassmann integrals, we shall impose a cutoff $M\in\mathbb N$: let $\chi_0(\rho)$ be a smooth compact support function that returns $1$ if $\rho\leqslant 1/3$ and $0$ if $\rho\geqslant 2/3$, and let
\begin{equation}
  \mathcal B_{\beta,L}^*:=\mathcal B_{\beta,L}\cap\{(k_0,k),\ \chi_0(2^{-M}|k_0|\neq0)\}
  .
\end{equation}
To every $\hat\alpha_{\mathbf k,\sigma}^{\pm}$ for $\alpha\in\{a,b\}$ and $\mathbf k\in\mathcal B_{\beta,L}^*$, we associate a {\it Grassmann variable} $\hat\psi_{\mathbf k,\alpha,\sigma}^\pm$, and we consider the finite Grassmann algebra (an algebra in which the $\hat\psi$ anti-commute with each other) generated by the collection $\{\hat\psi_{\mathbf k,\alpha,\sigma}^\pm\}_{\mathbf k\in\mathcal B_{\beta,L}^*}^{\alpha\in\{a,b\},\sigma\in\{\uparrow,\downarrow\}}$.
We define the Grassmann integral
\begin{equation}
  \int
  \prod_{\sigma\in\{\uparrow,\downarrow\}}\prod_{\alpha\in\{a,b\}}\prod_{\mathbf k\in{\mathcal B}^*_{\beta,L}}d\hat\psi_{\mathbf k,\alpha,\sigma}^+ d\hat\psi_{\mathbf k,\alpha,\sigma}^-
\end{equation}
as the linear operator on the Grassmann algebra whose action on a monomial in the variables $\hat\psi^\pm_{\mathbf k,\alpha,\sigma}$ is $0$ except if said monomial is $\prod_{\sigma\in\{\uparrow,\downarrow\}}\prod_{\alpha\in\{a,b\}}\prod_{\mathbf k\in{\mathcal B}_{\beta,L}^*} \hat\psi^-_{\mathbf k,\alpha,\sigma} \hat\psi^+_{\mathbf k,\alpha,\sigma}$ up to a permutation of the variables, in which case the value of the integral is determined using
\begin{equation}
  \int
  \prod_{\sigma\in\{\uparrow,\downarrow\}}\prod_{\alpha\in\{a,b\}}\prod_{\mathbf k\in{\mathcal B}_{\beta,L}^*}
  d\hat\psi_{\mathbf k,\alpha,\sigma}^+
  d\hat\psi_{\mathbf k,\alpha,\sigma}^-
  \left(\prod_{\sigma\in\{\uparrow,\downarrow\}}\prod_{\alpha\in\{a,b\}}\prod_{\mathbf k\in{\mathcal B}_{\beta,L}^*}
  \hat\psi^-_{\mathbf k,\alpha,\sigma}
  \hat\psi^+_{\mathbf k,\alpha,\sigma}\right)=1
\end{equation}
along with the anti-commutation of the $\hat\psi$.
\bigskip

\indent
Let us now define {\it Gaussian Grassmann integrals}.
The Gaussian Grassmann measure is specified by a {\it propagator}, which is a $2\times2$ complex matrix $\hat g(\mathbf k)$:
\begin{equation}
  \begin{largearray}
    P_{\hat g}(d\psi) := \left(
      \prod_{\mathbf k\in\mathcal B_{\beta,L}^*}
      (\beta^2\det\hat g(\mathbf k))^2
      \left(\prod_{\sigma\in\{\uparrow,\downarrow\}}\prod_{\alpha\in\{a,b\}}d\hat\psi_{\mathbf k,\alpha}^+d\hat\psi_{\mathbf k,\alpha}^-\right)
    \right)
    \cdot\\[0.5cm]\hfill\cdot
    \exp\left(-\frac{1}{\beta}\sum_{\sigma\in\{\uparrow,\downarrow\}}\sum_{\mathbf k\in\mathcal B_{\beta,L}^*}\hat\psi^{+}_{\mathbf k,\cdot,\sigma}\cdot\hat g^{-1}(\mathbf k)\hat\psi^{-}_{\mathbf k,\cdot,\sigma}\right)
    .
    \label{grassgauss}
  \end{largearray}
\end{equation}
By a direct computation, one can prove that (see lemma\-~\ref{lemma:grassmann_id})
\begin{equation}
  \int P_{\hat g}(d\psi)\ 1=1
  ,\quad
  \int P_{\hat g}(d\psi)\ \hat\psi_{\mathbf k,\alpha,\sigma}^-\hat\psi_{\bar\mathbf k,\bar\alpha,\bar\sigma}^+
  =\beta\delta_{\mathbf k,\bar{\mathbf k}}\delta_{\sigma,\bar\sigma}\hat g_{\alpha,\bar\alpha}(\mathbf k)
  .
  \label{twopt_grassmann}
\end{equation}
In addition, the integral with respect to $P_{\hat g}(d\psi)$ satisfies the Wick rule (see lemma\-~\ref{lemma:wick_grassmann}):
\begin{equation}
  \int P_{\hat g}(d\psi)
  \prod_{i=1}^{n}\hat\psi_{\mathbf k_i,\alpha_i,\sigma_i}^-\hat\psi_{\bar\mathbf k_i,\bar\alpha_i,\bar\sigma_i}^+
  =
  \sum_{\tau\in\mathcal S_n}(-1)^\tau
  \prod_{i=1}^{n}
  \int P_{\hat g}(d\psi)\ \hat\psi_{\mathbf k_i,\alpha_i,\sigma_i}^-\hat\psi_{\bar\mathbf k_{\tau(i)},\bar\alpha_{\tau(i)},\bar\sigma_{\tau(i)}}^+
  \label{wick_grassmann}
\end{equation}
Finally, given two propagators $\hat g_1$ and $\hat g_2$, and any polynomial $\mathfrak P(\psi)$ in the Grassmann variables, we have (see lemma\-~\ref{lemma:grassmann_addition}),
\begin{equation}
  \int P_{\hat g_1+\hat g_2}(d\psi)\ \mathfrak P(\psi)=\int P_{\hat g_1}(d\psi_1)\int P_{\hat g_2}(d\psi_2)\ \mathfrak P(\psi_1+\psi_2)
  .
  \label{addprop}
\end{equation}
\bigskip

\point{\bf Grassmann integrals and the free energy.}
Let us now make the connection between the computation of the free energy and Grassmann integrals.
As we have seen, free averages of polynomials in the creation and annihilation operators can be computed using the Wick rule, see\-~(\ref{wick}).
Gaussian Grassmann integrals of Grassmann polynomials also satisfy the Wick rule, see\-~(\ref{wick_grassmann}), so Gaussian Grassmann integrals and the free average share the same algebraic structure.
Thus, if we set $\hat g$ in such a way that
\begin{equation}
  \int P_{\hat g}(d\psi)\ \hat\psi_{\mathbf k,\alpha,\sigma}^-\hat\psi_{\bar\mathbf k,\bar\alpha,\bar\sigma}^+
  =
  \frac1{|\Lambda|}\sum_{x,\bar x\in\Lambda}\int_0^\beta dt\int_0^\beta d\bar t\ e^{itk_0-i\bar t\bar k_0+ikx-i\bar k\bar x}
  \left<\mathbf T\left(\alpha_{x,\sigma}^-(t)\bar\alpha^+_{\bar x,\bar\sigma}(\bar t)\right)\right>
\end{equation}
then we can compute free averages using Gaussian Grassmann integrals.
Furthermore, by a direct computation (see lemma\-~\ref{lemma:schwinger}),
\begin{equation}
  \frac1{|\Lambda|}\sum_{x,\bar x\in\Lambda}\int_0^\beta dt\int_0^\beta d\bar t\ e^{itk_0-i\bar t\bar k_0+ikx-i\bar k\bar x}
  \left<\mathbf T\left(\alpha_{x,\sigma}^-(t)\bar\alpha^+_{\bar x,\bar\sigma}(\bar t)\right)\right>
  =\beta\delta_{\mathbf k,\bar{\mathbf k}}\delta_{\sigma,\bar\sigma}(-ik_0\mathds 1-H_0(k))^{-1}
\end{equation}
so, by\-~(\ref{twopt_grassmann}),
\begin{equation}
  \hat g(\mathbf k)=(-ik_0\mathds 1-H_0(k))^{-1}
  .
\end{equation}
Actually, since we cut off the momenta by $M$, in order to avoid introducing Gibbs phenomena when inverting Fourier transforms, we define the {\it propagator}:
\begin{equation}
  \hat g_{\leqslant M}(\mathbf k):=\chi_0(2^{-M}|k_0|) (-ik_0\mathds 1-H_0(k))^{-1}
  \label{freeprop}
\end{equation}
and will take the limit $M\to\infty$ at the end of the computation.
Thus, we define the Gaussian Grassmann integration measure $P_{\leqslant M}(d\psi)\equiv P_{\hat g_{\leqslant M}}(d\psi)$, which allows us to compute the trace in\-~(\ref{freeen}):
\begin{equation}
  f_\Lambda=f_{0,\Lambda}-\lim_{M\to\infty}\frac{1}{\beta|\Lambda|}\log\int P_{\leqslant M}(d\psi)\ e^{-\mathcal V(\psi)}
  \label{freeengrass}
\end{equation}
where $f_{0,\Lambda}$ is the free energy in the $U=0$ case and
\begin{equation}
  \mathcal V(\psi)=U\sum_{\alpha\in\{a,b\}}\int_{0}^\beta dt \sum_{x\in \Lambda}\psi^+_{\mathbf x,\alpha,\uparrow}\psi^{-}_{\mathbf x,\alpha,\uparrow}
  \psi^+_{\mathbf x,\alpha,\downarrow}\psi^{-}_{\mathbf x,\alpha,\downarrow}
  \label{V_grassmann}
\end{equation}
in which $\mathbf x\equiv(t,x)$ and
\begin{equation}
  \psi^\pm_{\mathbf x,\alpha,\sigma}:=\frac1{\beta\sqrt{|\Lambda|}}\sum_{\mathbf k\in{\mathcal B}^*_{\beta,L}}\hat \psi^{\pm}_{\mathbf k,\alpha,\sigma}e^{\pm i\mathbf k\mathbf x}
  .
\end{equation}
\bigskip

\indent
Note that we have dropped the $-\frac12$ from the interaction Hamiltonian when passing to the Grassmann variables.
The reason for this is that, in configuration space, the point $t=\bar t$ is important (this situation did not arise above as this was of measure $0$).
Recall that in lemma\-~\ref{lemma:schwinger}, the two point function $\left<\mathbf T(\alpha^-\alpha^+)\right>$ at $t=\bar t$ has an extra $\frac12$, and one can show that this exactly cancels with the $-\frac12$ in the interaction Hamiltonian.

\subsection{Singularities of the propagator and scale decomposition}
\indent
We thus have a clear strategy to compute the free energy of the electrons in graphene: (\ref{freeengrass}) reduces the computation to a Gaussian Grassmann integral, which we can compute as a power series by expanding $e^{-\mathcal V}$.
However, when implementing this strategy, one almost immediately runs into a problem: $\hat g$ is singular, at least in the limit $\beta,L\to\infty$.
Indeed, recall that the eigenvalues of $H_0$ are singular at $k=p_F^{(\omega)}$, see\-~(\ref{fermipt}), around which they behave like $v_F|k-p_F^{(\omega)}|$, see\-~(\ref{dispersion}).
Thus, $\hat g_{\leqslant M}$ is singular at $\mathbf k=\mathbf p_F^{(\omega)}:=(0,p_F^{(\omega)})$, near which the eigenvalues of $\hat g_{\leqslant M}$ behave like
\begin{equation}
  \pm|v_F|k-p_F^{(\omega)}|+ik_0|^{-1}
  .
\end{equation}
(Note that $|k_0|\geqslant\frac\pi\beta$, so $\hat g$ is not singular for finite $\beta$, but it becomes so in the limit.)
Therefore, it is far from obvious that our strategy to compute the free energy will work uniformly in $\beta$, and so we may not be able to take the limit $\beta\to\infty$ (the risk is that the series expansion does not converge in the limit).
\bigskip

\begin{figure}
\hfil\includegraphics[width=5cm]{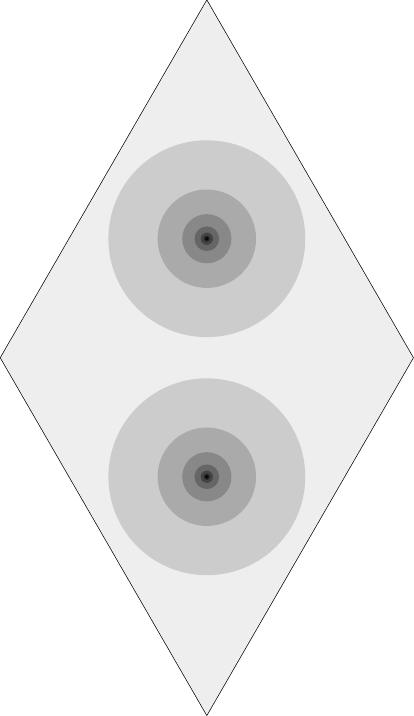}
\caption{
  A sketch of the scale decomposition, the darker the color, the smaller the scale.
}
\label{fig:scale}
\end{figure}

\indent
We will need to proceed in a more subtle fashion, by performing a {\it multiscale decomposition} (the multiscale decomposition is the heart of the renormalization group method).
The idea is to approach the singularities $p_F^{(\omega)}$ slowly, by defining scale-by-scale propagators: for $h\in-\mathbb N$, we define
\begin{equation}
  \Phi_{h}(\mathbf k-\mathbf p_F^{(\omega)}):=(\chi_0(2^{-h}|\mathbf k-\mathbf p_F^{(\omega)}|)-\chi_0(2^{-h+1}|\mathbf k-\mathbf p_F^{(\omega)}|))
  \label{fh}
\end{equation}
which is a smooth function that is supported in $|\mathbf k-\mathbf p_F^{(\omega)}|\in[2^h\frac16,2^h\frac23]$, in other words, it localizes $\mathbf k$ to be at a distance from $\mathbf p_F^{(\omega)}$ that is of order $2^h$, see figure\-~\ref{fig:scale}.
Since $|k_0|\geqslant\frac\pi\beta$, we only need to consider
\begin{equation}
  h\geqslant -N_\beta:=\log_2\frac\pi\beta
  .
\end{equation}
We then define
\begin{equation}
  \hat g^{(h,\omega)}(\mathbf k):=\Phi_h(\mathbf k-\mathbf p_F^{(\omega)})(-ik_0\mathds 1-H_0(k))^{-1}
  =\Phi_h(\mathbf k-\mathbf p_F^{(\omega)})
  \frac1{k_0^2+|\Omega(k)|^2}
  \left(\begin{array}{cc}
    ik_0&-\Omega^*(k)\\
    -\Omega(k)&ik_0
  \end{array}\right)
  \label{prop}
\end{equation}
(where we used\-~(\ref{hmat}) to compute the inverse).
For $h$ sufficiently small (sufficiently negative), we thus have
\begin{equation}
  \hat g^{(h,\omega)}=O(2^{-h}).
  \label{propscale}
\end{equation}
To carry out the Gaussian Grassmann integral, we split
\begin{equation}
  \hat g_{\leqslant M}=
  \sum_{h=0}^{-N_\beta}\sum_{\omega=\pm}\hat g^{(h,\omega)}(\mathbf k)
  +
  \hat g_{\geqslant 0}
  ,\quad
  \hat g_{\geqslant 0}:=
  \hat g_{\leqslant M}-\sum_{h=0}^{-N_\beta}\sum_{\omega=\pm}\hat g^{(h,\omega)}(\mathbf k)
\end{equation}
(note that $\hat g_{\geqslant 0}$ is bounded uniformly in $\beta$) and use\-~(\ref{addprop}) to compute the integrals with the propagators $g^{(h,\omega)}$ one at a time.
\bigskip

\bigskip

\indent
This has been done in detail in\-~\cite{GM10,Gi10}.
Carrying out this strategy in practice is rather involved though, so here, we will simplify the problem by considering a {\it hierarchical model} based on the graphene model.

\section{Hierarchical graphene}\label{sec:hierarchical_graphene}
\indent
The hierarchical graphene model is a simplification of the graphene model, in which everything but the multiscale structure is culled.
The philosophy of using a hierarchical model is that renormalization group treatments are typically quite involved, whereas hierarchical models behave in essentially (from a renormalization group point of view) the same way, without having too many details muddy the main ideas of the strategy.
They have been used in many settings\-~\cite{Dy69,BCe78,GK81}, though most of those are Bosonic.
Fermionic hierarchical models were first studied in\-~\cite{Do91}, and, in a more systematic way, in\-~\cite{BGJ15,GJ15}.
\bigskip

\subsection{Definition of the hierarchical model}
\begin{figure}
\hfil\includegraphics[width=12cm]{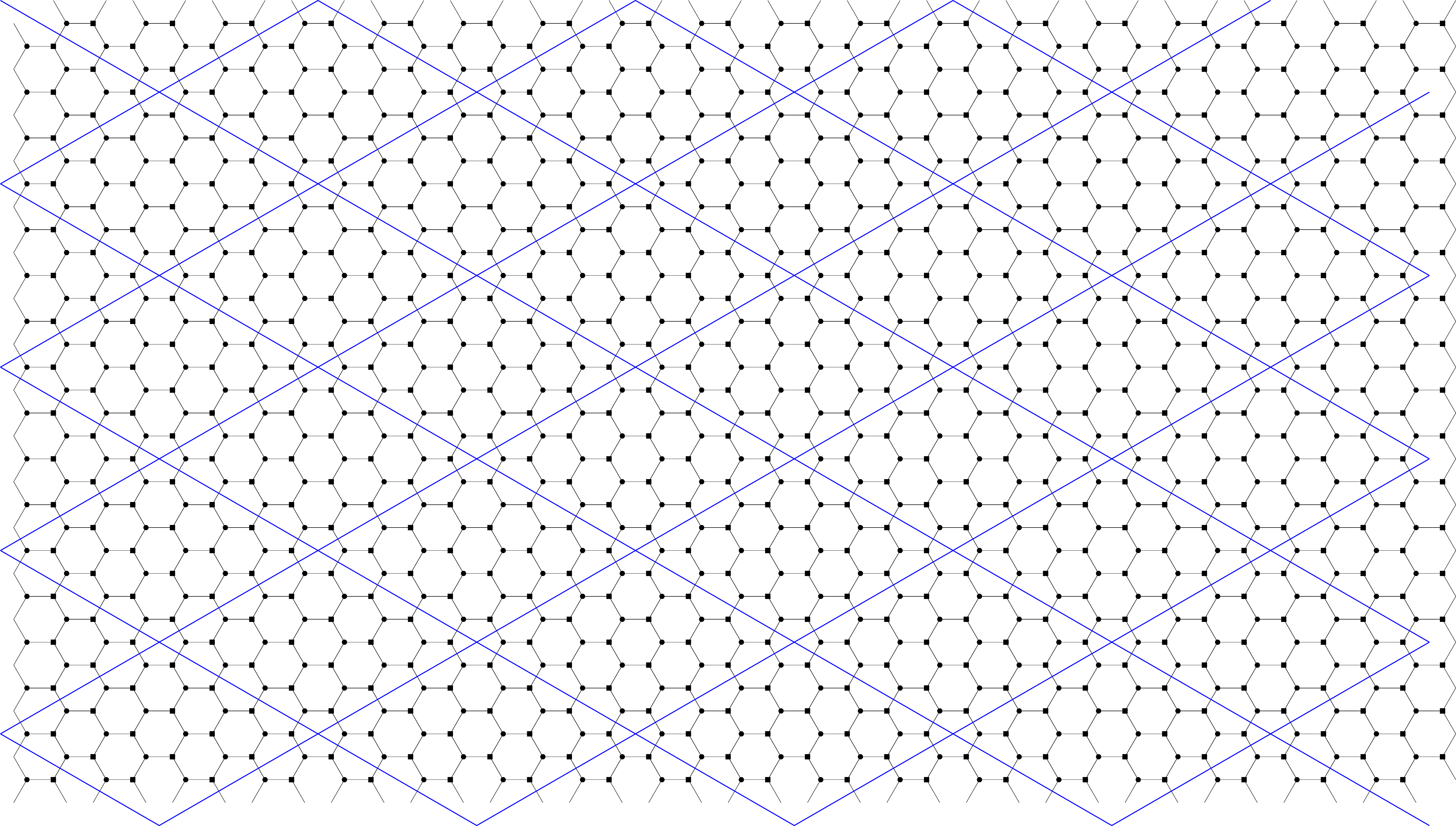}
\caption{
  In this figure, we have quadrupled the size of the elementary cell.
  Each large cell supports four Grassmann fields corresponding to both choices of type of atom ($a$ and $b$) and of spin ($\uparrow$ and $\downarrow$).
}
\label{fig:coarse}
\end{figure}

\point{\bf Boxes.}
The hierarchical model will be defined in configuration space (as opposed to Fourier space).
In the previous section, we introduced a multiscale decomposition in Fourier space, in which we split Grassmann fields into momentum shells, where $|\mathbf k-\mathbf p_F^{(\omega)}|\sim2^h$.
In configuration space, this corresponds to considering Grassmann fields that are constant on configuration-space boxes of size $\sim 2^{-h}$.
These boxes are obtained by doubling the size of the elementary cell of the hexagonal lattice at every scale, see figure\-~\ref{fig:coarse}.
We also have a time dimension, which we split into boxes of size $2^{|h|}$.
We thus define the set of boxes on scale $h\in\{-N_\beta,\cdots,0\}$ by
\begin{equation}
  \mathcal Q_h:=\left\{
    [i2^{|h|},(i+1)2^{|h|})\times(\Lambda\cap\{2^{|h|}(n_1+x_1)l_1+2^{|h|}(n_2+x_2)l_2,\ x_1,x_2\in[0,1)\})
  \right\}_{i,n_1,n_2\in\mathbb Z}
  \label{boxes}
\end{equation}
($\mathcal Q_h$ is a set of sets).
For every $(t,x)\in[0,\beta)\times\Lambda$ and $h\in\{-N_\beta,\cdots,0\}$, there exists a unique box $\Delta^{(h)}(t,x)\in\mathcal Q_m$ such that $(t,x)\in\Delta^{(m)}(t,x)$.
To simplify the computations further, we drop the index $\omega$, which does not change the nature of the problem.
In each box $\Delta$, we have four Grassmann fields (and their conjugates), corresponding to both choices of atom type ($a$ or $b$) and both choices of spin ($\uparrow$ and $\downarrow$): for $\alpha\in\{a,b\}$ and $\sigma\in\{\uparrow,\downarrow\}$, $\psi_{\alpha,\sigma}^{[h]\pm}(\Delta)$.
\bigskip

\bigskip

\point{\bf Propagators.}
Let us now define the propagators associated to these Grassmann fields.
We will chose them to be {\it similar}, in a certain sense, to the non-hierarchical propagators.
We defined the single scale propagator in Fourier space in\-~(\ref{prop}), which, in configuration space, becomes
\begin{equation}
  g^{(h,\omega)}(t,x)=
  \frac1{\beta|\Lambda|}\sum_{\mathbf k\in\mathcal B_{\beta,L}}e^{-ik_0t-ikx}
  \Phi_h(\mathbf k-\mathbf p_F^{(\omega)})
  \frac1{k_0^2+|\Omega(k)|^2}
  \left(\begin{array}{cc}
    ik_0&-\Omega^*(k)\\
    -\Omega(k)&ik_0
  \end{array}\right)
  .
\end{equation}
In the hierarchical approximation, we neglect propagators at different times, so we can set $t=0$, at which, by symmetry, the diagonal terms vanish.
Thus we only have a non-trivial propagator in between atoms of type $a$ and $b$.
Furthermore,
\begin{equation}
  \frac1{\beta|\Lambda|}\sum_{\mathbf k\in\mathcal B_{\beta,L}}
  \Phi_h(\mathbf k-\mathbf p_F^{(\omega)})
  =O(2^{3h})
  ,\quad
  \hat g^{(h,\omega)}=O(2^{-h})
\end{equation}
so
\begin{equation}
  g^{(h,\omega)}(t,x)=O(2^{2h})
  .
  \label{propx}
\end{equation}
In order to take this scaling factor into account, we rescale the Grassmann fields $\psi$, and define, for $(t,x)\in[0,\beta)\times\Lambda$, (recall that we have dropped the index $\omega$ in the hierarchical model)
\begin{equation}
  \psi_{\alpha,\sigma}^\pm(t,x)=\sum_{h=-N_\beta}^02^h\psi_{\alpha,\sigma}^{[h]\pm}(\Delta^{(h)}(t,x))
  .
  \label{psi_hierarchical}
\end{equation}
We will take the propagators to be
\begin{equation}
  \int P^{[h]}(d\psi^{[h]})\ \psi_{a,\sigma}^{[h]-}(\Delta)\psi_{b,\sigma'}^{[h]+}(\Delta')=\delta_{\sigma,\sigma'}\delta_{\Delta,\Delta'}
\end{equation}
\begin{equation}
  \int P^{[h]}(d\psi^{[h]})\ \psi_{b,\sigma}^{[h]-}(\Delta)\psi_{a,\sigma'}^{[h]+}(\Delta')=\delta_{\sigma,\sigma'}\delta_{\Delta,\Delta'}
\end{equation}
and all other propagators will be set to 0.
We can now evaluate how well these propagators approximate the non-hierarchical ones.
Given $x,y$, let $\eta(x,y)$ be the largest negative integer such that $x$ and $y$ are in the same box.
We have
\begin{equation}
  \int P(d\psi)\ \psi_{a,\sigma}^-(x,t)\psi_{b,\sigma'}^+(y,t)
  =
  \delta_{\sigma,\sigma'}
  \sum_{h=-N_\beta}^{\eta(x,y)}
  2^{2h}
  =
  \delta_{\sigma,\sigma'}
  \frac432^{2\eta(x,y)}(1-4^{-N_\beta-\eta(x,y)})
\end{equation}
which has the same scaling as the non-hierarchical propagator\-~(\ref{propx}) (and similarly with $a$ and $b$ exchanged).
\bigskip

\point{\bf Effective potentials.}
In this hierarchical model, we compute
\begin{equation}
  \int P(d\psi)\ e^{-\mathcal V(\psi)}
  .
\end{equation}
We proceed inductively: for $\alpha\in\{a,b\}$, $\sigma\in\{\uparrow,\downarrow\}$, $h\in\{-N_\beta,\cdots,0\}$, $\Delta\in\mathcal Q_h$, let $\bar\Delta\in\mathcal Q_{h-1}$ be such that $\bar\Delta\supset\Delta$, we define
\begin{equation}
  \psi_{\alpha,\sigma}^{[\leqslant h]\pm}(\Delta)
  :=
  \frac12\psi_{\alpha,\sigma}^{[\leqslant h-1]\pm}(\bar\Delta)
  +
  \psi_{\alpha,\sigma}^{[h]\pm}(\Delta)
  .
  \label{scaling_psi}
\end{equation}
Note that\-~(\ref{psi_hierarchical}) is then
\begin{equation}
  \psi_{\alpha,\sigma}^\pm(t,x)\equiv\psi_{\alpha,\sigma}^{[\le0]\pm}(\Delta^{(1)}(t,x))
  .
\end{equation}
We then define, for $h\in\{-N_\beta,\cdots0\}$,
\begin{equation}
  e^{\beta|\Lambda| c^{[h]}-\mathcal V^{[h-1]}(\psi^{\leqslant h-1]})}
  :=\int P^{[h]}(d\psi^{[h]})\ e^{-\mathcal V^{[h]}(\psi^{[\leqslant h]})}
  ,\quad
  \mathcal V^{[0]}(\psi^{[\le0]}):=\mathcal V(\psi^{[\le0]})
  \label{indV}
\end{equation}
in which $c^{[h]}\in\mathbb R$ is a constant and $\mathcal V^{[h-1]}$ has no constant term.
The function $\mathcal V^{[h]}$ is called the {\it effective potential} on scale $h$, and it dictates the physical properties of the system at distances $\sim 2^{-h}$.
By a straightforward induction, we then find that
\begin{equation}
  \int P(d\psi)\ e^{-\mathcal V(\psi)}
  =\exp\left({\textstyle-\beta|\Lambda|\sum_{h=-N_\beta}^0c^{[h]}}\right)
  .
\end{equation}
The specific free energy\-~(\ref{freeengrass}) is then
\begin{equation}
  f_\Lambda=f_{0,\Lambda}-\sum_{h=-N_\beta}^0c^{[h]}
  .
\end{equation}
We are then left with computing $\mathcal V^{[h]}$ and $c^{[h]}$ using\-~(\ref{indV}).
By\-~(\ref{V_grassmann}), $\mathcal V$ is local in $(t,x)$ and so, by induction, it takes the form
\begin{equation}
  \mathcal V^{[h]}(\psi^{[\leqslant h]})=-\sum_{\Delta\in\mathcal Q_h}v_h(\psi^{[\leqslant h]}(\Delta))
  .
  \label{box_dcmp}
\end{equation}
\bigskip

\indent
Because there are only four Grassmann fields and their conjugates per cell, $v_h$ must be a {\it polynomial} in the Grassmann fields of order $\leqslant 8$.
In fact, by symmetry considerations, we find that $v_h$ must be of the form
\begin{equation}
  v_h(\psi)=
  \sum_{i=0}^6\ell_i^{(h)}O_i(\psi)
  \label{vh_rcc}
\end{equation}
with
\begin{equation}
  O_0(\psi):=
  \sum_{\sigma\in\{\uparrow,\downarrow\}}\left(
    \psi_{a,\sigma}^{+}
    \psi_{b,\sigma}^{-}
    +
    \psi_{b,\sigma}^{+}
    \psi_{a,\sigma}^{-}
  \right)
  ,\quad
  O_1(\psi):=
  \sum_{\alpha\in\{a,b\}}
  \psi_{\alpha,\uparrow}^{+}
  \psi_{\alpha,\uparrow}^{-}
  \psi_{\alpha,\downarrow}^{+}
  \psi_{\alpha,\downarrow}^{-}
\end{equation}
\begin{equation}
  O_2(\psi):=
  \psi_{a,\uparrow}^{+}
  \psi_{a,\downarrow}^{-}
  \psi_{b,\downarrow}^{+}
  \psi_{b,\uparrow}^{-}
  +
  \psi_{b,\uparrow}^{+}
  \psi_{b,\downarrow}^{-}
  \psi_{a,\downarrow}^{+}
  \psi_{a,\uparrow}^{-}
  +
  \psi_{a,\downarrow}^{+}
  \psi_{a,\uparrow}^{-}
  \psi_{b,\uparrow}^{+}
  \psi_{b,\downarrow}^{-}
  +
  \psi_{b,\downarrow}^{+}
  \psi_{b,\uparrow}^{-}
  \psi_{a,\uparrow}^{+}
  \psi_{a,\downarrow}^{-}
\end{equation}
\begin{equation}
  O_3(\psi):=
  \sum_{\sigma\in\{\uparrow,\downarrow\}}
  \psi_{a,\sigma}^+
  \psi_{a,\sigma}^-
  \psi_{b,\sigma}^+
  \psi_{b,\sigma}^-
  ,\quad
  O_4(\psi):=
  \psi_{a,\uparrow}^+
  \psi_{b,\uparrow}^-
  \psi_{a,\downarrow}^+
  \psi_{b,\downarrow}^-
  +
  \psi_{b,\uparrow}^+
  \psi_{a,\uparrow}^-
  \psi_{b,\downarrow}^+
  \psi_{a,\downarrow}^-
\end{equation}
\begin{equation}
  \begin{largearray}
    O_5(\psi):=
    \psi_{a,\uparrow}^+
    \psi_{a,\uparrow}^-
    \psi_{a,\downarrow}^+
    \psi_{b,\uparrow}^-
    \psi_{b,\uparrow}^+
    \psi_{b,\downarrow}^-
    +
    \psi_{a,\downarrow}^+
    \psi_{a,\downarrow}^-
    \psi_{a,\uparrow}^+
    \psi_{b,\downarrow}^-
    \psi_{b,\downarrow}^+
    \psi_{b,\uparrow}^-
    +\\[0.3cm]\hfill+
    \psi_{b,\uparrow}^+
    \psi_{b,\uparrow}^-
    \psi_{b,\downarrow}^+
    \psi_{a,\uparrow}^-
    \psi_{a,\uparrow}^+
    \psi_{a,\downarrow}^-
    +
    \psi_{b,\downarrow}^+
    \psi_{b,\downarrow}^-
    \psi_{b,\uparrow}^+
    \psi_{a,\downarrow}^-
    \psi_{a,\downarrow}^+
    \psi_{a,\uparrow}^-
  \end{largearray}
\end{equation}
\begin{equation}
  O_6(\psi):=
  \psi_{a,\uparrow}^+
  \psi_{a,\uparrow}^-
  \psi_{a,\downarrow}^+
  \psi_{a,\downarrow}^-
  \psi_{b,\uparrow}^+
  \psi_{b,\uparrow}^-
  \psi_{b,\downarrow}^+
  \psi_{b,\downarrow}^-
  .
\end{equation}

\subsection{Beta function of the hierarchical model}
\point{\bf Beta function.}
We have thus introduced a strategy to compute $\mathcal V^{[h]}$ inductively: starting from $\mathcal V^{[h]}$, we have
\begin{equation}
  e^{\beta|\Lambda|c^{[h]}-\mathcal V^{[h-1]}(\psi^{[\leqslant h-1]})}=\int P(d\psi^{[h]})\ e^{-\mathcal V^{[h]}(\psi^{[h]}+2^{-\gamma}\psi^{[\leqslant h-1]})}
\end{equation}
where $\gamma\equiv1$ is the scaling dimension of $\psi$ in\-~(\ref{scaling_psi}).
Now, by\-~(\ref{box_dcmp}), is
\begin{equation}
  e^{\beta|\Lambda|c^{[h]}+\sum_{\bar\Delta\in\mathcal Q_{h-1}}v_{h-1}(\psi^{[\leqslant h-1]}(\bar\Delta))}=
  \prod_{\Delta\in\mathcal Q_h}\int P(d\psi^{[h]}(\Delta))\ e^{v_h(\psi^{[h]}(\Delta)+2^{-\gamma}\psi^{[\leqslant h-1]}(\bar\Delta))}
  .
\end{equation}
If we group the right side in boxes on scale $h-1$, we find
\begin{equation}
  e^{\beta|\Lambda|c^{[h]}+\sum_{\bar\Delta\in\mathcal Q_{h-1}}v_{h-1}(\psi^{[\leqslant h-1]}(\bar\Delta))}=
  \prod_{\bar\Delta\in\mathcal Q_{h-1}}
  \prod_{\displaystyle\mathop{\scriptstyle\Delta\in\mathcal Q_h}_{\Delta\subset\bar\Delta}}
  \int P(d\psi^{[h]}(\Delta))\ e^{v_h(\psi^{[h]}(\Delta)+2^{-\gamma}\psi^{[\leqslant h-1]}(\bar\Delta))}
  .
\end{equation}
In addition, the integral over $\psi^{[h]}(\Delta)$ does not depend on $\Delta$.
Therefore, using the fact that $\bar\Delta$ contains $2^{d+1}\equiv 8$ ($d\equiv2$ is the dimension of the lattice) boxes $\Delta$, we have, for all $\bar\Delta\in\mathcal Q_{h-1}$ and for any choice of $\Delta\in\mathcal Q_h$ with $\Delta\subset\bar\Delta$,
\begin{equation}
  e^{\beta|\Lambda|c^{[h]}+v_{h-1}(\psi^{[\leqslant h-1]}(\bar\Delta))}=
  \left(\int P(d\psi^{[h]}(\Delta))\ e^{v_h(\psi^{[h]}(\Delta)+2^{-\gamma}\psi^{[\leqslant h-1]}(\bar\Delta))}\right)^{2^{d+1}}
  .
\end{equation}
We expand the exponential and use\-~(\ref{vh_rcc}):
\begin{equation}
  \begin{largearray}
    \beta|\Lambda|c^{[h]}
    +
    \sum_{i=0}^6\ell_i^{(h-1)}O_i(\psi^{[\leqslant h-1]}(\bar\Delta))
    =\\\hfill=
    2^{d+1}\log
    \int P(d\psi^{[h]}(\Delta))
    \sum_{n=0}^\infty
    \frac1{n!}
    \left(\sum_{i=0}^6\ell_i^{(h)}O_i\left(\psi^{[h]}(\Delta)+2^{-\gamma}\psi^{[\leqslant h-1]}(\bar\Delta)\right)\right)^n
    .
  \end{largearray}
  \label{betadef}
\end{equation}
The computation is thus reduced to computing the map $\ell^{(h)}\mapsto\ell^{(h-1)}$ using\-~(\ref{betadef}).
The coefficients $\ell_i^{(h)}$ are called {\it running coupling constants}, and the map $\ell^{(h)}\mapsto\ell^{(h-1)}$ is called the {\it beta function} of the model.
The running coupling constants play a very important role, as they specify the effective potential on scale $h$, and thereby the physical properties of the system at distances $\sim2^{-h}$.
\bigskip

\point{\bf Feynman diagrams.}
Having defined the hierarchical model as we have, the infinite sum in\-~(\ref{betadef}) is actually finite ($n\leqslant 4$), so to compute the beta function, it suffices to compute Gaussian Grassmann integrals of a finite number of Grassmann monomials.
A convenient way to carry out this computation is to represent each term graphically, using {\it Feynman diagrams}.
First, let us expand the power $n$ and graphically represent the terms that must be integrated.
For each $n$, we have $n$ possible choices of $\ell_iO_i$.
Now, $O_i$ can be quadratic in $\psi$ ($O_0$), quartic ($O_1$, $O_2$, $O_3$, $O_4$), sextic ($O_5$) or octic ($O_6$).
We will represent $O_i$ by a vertex with the label $i$, from which two, four, six or eight edges emanate, depending on the degree of $O_i$.
Each edge corresponds to a factor $\psi^{[h]}+2^{-\gamma}\psi^{[\leqslant h-1]}$.
For each edge, we can either choose the term $2^{-\gamma}\psi^{[\leqslant h-1]}$, which is not integrated, and so the edge will be called {\it external}, or we can choose the term $\psi^{[h]}$, in which case this edge will have to be integrated, and will be called {\it internal}.
Having made all of these choices, we have a set of vertices as well as external and internal edges.
We must now integrate $\psi^{[h]}$.
Recall that, by Wick's rule\-~(\ref{wick_grassmann}), to integrate a monomial, we pair up $\psi^{[h]}$'s (internal edges), and multiply the corresponding propagators.
Graphically, this is done by connecting internal edges.
Each connection made stands in for a propagator.
The computation of the Gaussian Grassmann integral in\-~(\ref{betadef}) can thus be reduced to a sum over graphs.
\bigskip

\indent
Let us make a few comments about this expansion, and how it is used for non-hierarchical models.

\begin{itemize}
  \item
  The diagram expansion for the hierarchical model is finite, in that there are only finitely many types of vertices, and finitely many of them, so the number of possible graphs is finite.
  This diagrammatic representation can be done in the non-hierarchical model as well, but there, the number of graphs is infinite.
  In fact, even considering just one type of vertex, the number of graphs grows as $(n!)^2$ as the number of vertices $n$ tends to $\infty$.
  This makes the infinite sum in\-~(\ref{betadef}) absolutely divergent.
  However, in counting these graphs, we are ignoring important signs: in the Wick rule\-~(\ref{wick_grassmann}), each permutation comes with its signature.
  These signs lead to important cancellations in the sum over $n$, which can be exploited using so-called {\it Gram bounds}, or {\it determinant expansions}\-~\cite{BF78,BF84} (these are even important numerically, as they can be used to compute the same integrals with many fewer terms).
  Here, we will focus on the hierarchical model for which such considerations are not needed.
  The Gram bounds for graphene are worked out in detail in\-~\cite[Appendix\-~B]{Gi10}.

  \item
  The diagram expansion allows us to compute the integral in\-~(\ref{betadef}).
  One then has to take the logarithm, which can also be done by an expansion.
  The sign in the expansion of the logarithm leads to significant cancellations, which, it turns out, cancels all of the graphs that are not connected.
  We will not dwell on this fact here, as this is not so relevant for the hierarchical model, but this is an important fact for non-hierarchical models.

  \item
  The diagram expansion we have discussed performs the integral on a single scale.
  In the grander scheme, the external edges will get contracted on lower scales, and one can construct a larger diagram expansion that covers all scales.
  Understanding the combinatorial properties of this full expansion requires some extra tools to keep track of the scales of the edges.
  One way of doing this is to use Gallavotti-Nicol\'o\-~\cite{GN85} trees, as detailed in\-~\cite{Gi10}.
  Another good reference for the tree expansion and its connection to Feynman diagrams is\-~\cite[Section\-~5]{GJ16}
  This is not needed for the hierarchical model.
\end{itemize}
\bigskip

\point{\bf Power counting.}
Returning to our hierarchical model, let us consider some of the more important contributions to the Feynman diagram expansion: bare vertices.
These are the simplest possible graphs, in which we have one vertex, and all edges are external.
In other words, no integrating is taking place.
Let us denote the number of external edges by $2l$, which can either be 2, 4, 6 or 8.
The contribution of this graph is (keeping track of the $2^{d+1}$ factor in\-~(\ref{betadef}))
\begin{equation}
  2^{d+1-2l\gamma}\ell_i^{(h)}
  .
\end{equation}
Furthermore, this graph will contribute to the running coupling constant $\ell_i$, and so, on scale $h-1$, we will have
\begin{equation}
  \ell_i^{(h-1)}=
  2^{d+1-2l\gamma}\ell_i^{(h)}
  +
  \cdots
\end{equation}
(the $\cdots$ stand in for the other terms contributing the this running coupling constant, and will be discarded for the moment).
Three things can happen:
\begin{itemize}
  \item
  if $d+1-2l\gamma<0$, then this term will decrease exponentially as $h\to-\infty$; such running coupling constants will be called {\it irrelevant};
  \item
  if $d+1-2l\gamma>0$, then this term will grow exponentially as $h\to-\infty$; such running coupling constants will be called {\it relevant};
  \item
  if $d+1-2l\gamma=0$, then not much can be said without looking at the neglected terms; such running coupling constants will be called {\it marginal}.
\end{itemize}
In the case of graphene, $d=2$ and $\gamma=1$, so if $2l=2$, then the running coupling constant is relevant, and if $2l\geqslant 4$, it is irrelevant; there are no marginal constants.
\bigskip

\indent
But what of the other terms we neglected?
The notions of relevant, irrelevant and marginal terms only apply to a perturbative setting, in which the running coupling constants are small, and the question is whether they stay small.
If all running coupling constants are irrelevant, then if they start small at scale $h=0$, they stay that way for all scales.
When some are relevant or marginal, there is a risk for some to grow and leave the perturbative regime.
Since the running coupling constants dictate the physical properties of the system, leaving the perturbative regime means that, on the scales on which the constants grow, the system will start behaving radically differently from the non-interacting one (in which all running coupling constants are 0).
In other words, studying whether running coupling constants are relevant, irrelevant or marginal, gives us information about whether the interacting system can be approximated by the non-interacting one or not.
\bigskip

\indent
This {\it power counting} holds for the non-hierarchical model as well, as detailed in\-~\cite{Gi10}.
For a more general treatment of power counting in Fermionic models with point-singularities, see\-~\cite[Section\-~5.2]{GJ16}.
\bigskip

\indent
In the case of graphene, we have one relevant coupling: $O_0$, which is quadratic in the Grassmann fields.
This is the only relevant coupling, and all others stay small.
However, since the relevant coupling is quadratic, it merely shifts the non-interacting system (whose Hamiltonian is quadratic in the Grassmann fields) to another system with a quadratic (that is, non-interacting) Hamiltonian.
Thus the relevant coupling does {\it not} imply that the interactions are preponderant, but rather that the interaction terms shifts the system from one non-interacting system to another.
Since graphene only has one relevant coupling, and that one is quadratic, graphene is called {\it super-renormalizable}.
\bigskip

\point{\bf Hierarchical beta function.}
As was mentioned above, the beta function can be computed {\it explicitly} for the hierarchical model, so the claims in the previous paragraph can be verified rather easily.
The exact computation involves many terms, but it can be done easily using the {\tt meankondo} software package\-~\cite{mk}.
The resulting beta function contains 888 terms, and will not be written out here.
A careful analysis of the beta function shows that there is an equilibrium point at $\ell_i=0$ for $i=1,2,3,4,5,6$ and
\begin{equation}
  \ell_0\in\{0,1\}
  .
\end{equation}
The point with $\ell_0=0$ is unstable, whereas $\ell_0=1$ is stable.
\bigskip

\begin{figure}
\hfil\includegraphics[width=12cm]{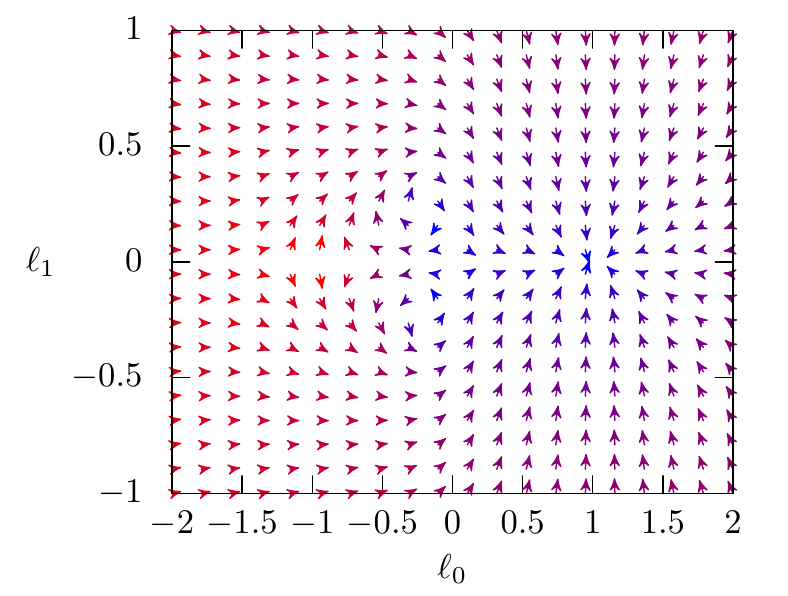}
\caption{
  The projection of the directional vector field of the beta function for hierarchical graphene onto the $(\ell_0,\ell_1)$ plane.
  (Each arrow shows the direction of the vector field, the color corresponds to the logarithm of the amplitude, with red being larger and blue smaller.)
  The stable equilibrium point at $\ell_0=1$ and $\ell_i=0$ is clearly visible.
}
\label{fig:vector_field}
\end{figure}

\subsection{Results for non-hierarchical graphene}
\indent
Using the Gram bounds and Gallavotti-Nicol\'o tree expansion hinted at above, the renormalization group analysis can be carried out in full for (non-hierarchical) graphene\-~\cite{Gi10,GM10}.
The result is quite similar to what we have found for the hierarchical model: the effective potential contains only irrelevant terms, except for the quadratic terms, which one absorbs into the quadratic non-interacting Hamiltonian.
We thus find that the observables for the interacting model are similar to those without interactions, provided the free Hamiltonian is suitably changed (which is called {\it renormalizing} the non-interacting Hamiltonian).
In particular, the Fermi velocity $v_F$ in\-~(\ref{dispersion}) is renormalized.
\bigskip

\indent
This renormalization can also be quantified: the renormalization group analysis provides a power series expansion in $U$, and, when using the Gram bounds, this series is shown to be absolutely convergent.
(In other words, the observables are analytic in $U$.)
Thus, we can estimate the effect of the interaction by truncating the power series and bounding its remainder.
\bigskip

\indent
In the case of the non-hierarchical graphene model, the renormalization of the free Hamiltonian is restricted by symmetries.
This is a crucial ingredient of the construction.
Indeed, while iterating the renormalization group flow, the free Hamiltonian is changed.
However, the philosophy of the scale decomposition is anchored in the singularities of the (inverse of the) non-interacting Hamiltonian, so if these singularities were to change under the flow, the entire theory might break down.
The symmetries prevent that from happening: in particular, they ensure that the singularities $p_F^\omega$ remain point singularities (they could a priori, turn them into curves).
This is crucial, as extended singularities would change the power counting, and would not allow the renormalization group flow to extend to arbitrarily small scales.
(The symmetries also ensure that the singularities do not move, but if they did, this could be dealt with by defining the scale decomposition slightly differently, see\-~\cite{GJ16} for an analysis of bilayer graphene where this happens).

\section{Hierarchical Kondo model}\label{sec:hierarchical_kondo}
\indent
The graphene model discussed above is {\it perturbative}, in the sense that the interactions do not qualitatively change the behavior of the system.
Let us now discuss a different model where the interactions change the behavior of the system drastically: the Kondo model.
The Kondo model, first studied by Kondo\-~\cite{Ko64}, was one of the foundational models in the development stages of the renormalization group\-~\cite{Wi75}.
It is integrable\-~\cite{An80}, but is nevertheless an interesting model to study renormalization group flows in.
The hierarchical Kondo model was introduced and studied in\-~\cite{BGJ15,GJ15}.
\bigskip

\subsection{Hamiltonian}
\indent
The Kondo model is a one-dimensional model of spin-$\frac12$ electrons on a chain that interact with a localized magnetic impurity, which is represented as a spin-$\frac12$ particle located at $x=0$.
The Hilbert space for the electrons is the usual Fock space $\mathcal F$, and the Hilbert space for the impurity is $\mathbb C^2$.
The Hamiltonian is split into the free Hamiltonian and the interaction term:
\begin{equation}
  \bar{\mathcal H}_0=\bar{\mathcal H}_0+\bar{\mathcal H_I}
\end{equation}
with
\begin{equation}
  \bar{\mathcal H}_0=-\frac12\sum_{\sigma\in\{\uparrow,\downarrow\}}\sum_{x\in\{-\frac L2+1,\cdots,\frac L2\}}a_{x,\sigma}^\dagger(a_{x+1,\sigma}+a_{x-1,\sigma})
  \otimes\mathds 1
\end{equation}
and
\begin{equation}
  \bar{\mathcal H}_I=-U\sum_{\sigma_1,\sigma_2\in\{\uparrow,\downarrow\}}\sum_{j=1,2,3}
  a_{0,\sigma_1}^\dagger S^{(j)}_{\sigma_1,\sigma_2}a_{0,\sigma_2}
  \otimes S^{(j)}
\end{equation}
where $a_{x,\sigma}$ is the annihilation operator at $x$ with spin $\sigma$ and $S^{(j)}$ are Pauli matrices:
\begin{equation}
  S^{(1)}=\left(\begin{array}{cc}
    0&1\\
    1&0
  \end{array}\right)
  ,\quad
  S^{(2)}=\left(\begin{array}{cc}
    0&-i\\
    i&0
  \end{array}\right)
  ,\quad
  S^{(3)}=\left(\begin{array}{cc}
    1&0\\
    0&-1
  \end{array}\right)
  .
\end{equation}
In Fourier space, let
\begin{equation}
  \hat\Lambda_L:=\frac{2\pi}L\left\{-\frac L2+1,\cdots,\frac L2\right\}
\end{equation}
in term of which
\begin{equation}
  \bar{\mathcal H}_0=-\sum_{\sigma\in\{\uparrow,\downarrow\}}\frac1L\sum_{k\in\hat\Lambda_L}\cos(k)\hat a_{k,\sigma}^\dagger\hat a_{k,\sigma}
  \otimes\mathds 1
  .
\end{equation}
The propagator is thus
\begin{equation}
  g(t,x)=\frac1{\beta L}\sum_{k_0\in\frac{2\pi}\beta(\mathbb Z+\frac12)}\sum_{k\in\hat\Lambda_L}\ e^{-ik_0t}e^{-ikx}\frac1{-ik_0-\cos(k)}
  .
\end{equation}

\subsection{Scale decomposition and hierarchical model}
\indent
Since the interaction is localized at $x=0$, we only need propagators at $x=0$:
\begin{equation}
  g(t,0)=\frac1{\beta L}\sum_{k_0\in\frac{2\pi}\beta(\mathbb Z+\frac12)}\sum_{k\in\hat\Lambda_L}\ e^{-ik_0t}\frac1{-ik_0-\cos(k)}
  .
\end{equation}
We decompose the propagator into scales around the singularities $(k_0,k)=(0,\pm\frac\pi2)$:
\begin{equation}
  g^{(h,\omega)}(t,0)=\frac1{\beta L}\sum_{k_0\in\frac{2\pi}\beta(\mathbb Z+\frac12)}\sum_{k\in\hat\Lambda_L}\ e^{-ik_0t}\frac1{-ik_0-\cos(k)}\Phi_h(k_0,k)
\end{equation}
where $\Phi_h$ is defined as in\-~(\ref{fh})
\begin{equation}
  \Phi_{h}(k_0,k):=(\chi_0(2^{-h}|(k_0,k-\omega{\textstyle\frac\pi2})|)-\chi_0(2^{-h+1}|(k_0,k-\omega{\textstyle\frac\pi2})|))
  .
\end{equation}
On scale $h$ we have
\begin{equation}
  \frac1{|ik_0+\cos(k)|}=O(2^{-h})
  ,\quad
  \sum_{k_0,k}\Phi_h(k_0,k)
  =O(2^{2h})
\end{equation}
so
\begin{equation}
  g^{(h,\omega)}(t,0)=O(2^h)
  \label{scaling_kondo}
\end{equation}
(as opposed to $2^{2h}$ for graphene).
\bigskip

\indent
We define the hierarchical model in a very similar way to graphene, except that, since only the point $x=0$ matters, there is no spatial dependence.
We define the boxes as
\begin{equation}
  \mathcal Q_h:=\left\{
    [i2^{|h|},(i+1)2^{|h|})
  \right\}_{i\in\mathbb Z}
  .
\end{equation}
We further split each box into two halves, which is necessary otherwise the model would come out trivial: for each box $\Delta=[i2^{|h|},(i+1)2^{|h|+1})$, we define
\begin{equation}
  \Delta_-=[i2^{|h|},(i+{\textstyle\frac12})2^{|h|})
  ,\quad
  \Delta_+=[(i+{\textstyle\frac12})2^{|h|},(i+1)2^{|h|})
\end{equation}
In accordance with the scaling in\-~(\ref{scaling_kondo}), we define, for $\sigma\in\{\uparrow,\downarrow\}$, $h\in\{-N_\beta,\cdots,-1\}$, $\Delta\in\mathcal Q_h$, $\eta\in\pm$
\begin{equation}
  \psi_{\sigma}^{[\leqslant h]\pm}(\Delta_\eta)
  :=
  \frac1{\sqrt2}\psi^{[\leqslant h-1]\pm}(\Delta)+\psi_{\sigma}^{[h]\pm}(\Delta_\eta)
\end{equation}
(compare this to\-~(\ref{scaling_psi}) in which the scaling factor is $\frac12$ instead of $\frac1{\sqrt2}$).

\subsection{Beta function}
\indent
For this model, it is more convenient to expand the exponential of the potential: we denote
\begin{equation}
  e^{-\mathcal V^{[h]}(\psi)}=\mathcal W^{[h]}(\psi)
  .
\end{equation}
One can show that the effective potentials $\mathcal W$ are of the form
\begin{equation}
  \mathcal W^{[h]}(\psi^{[\leqslant h]})
  =
  \prod_{\Delta\in\mathcal Q_h}\prod_{\eta=\pm}(1+w_h(\psi^{[\leqslant h]}(\Delta_\eta)))
  .
\end{equation}
For this model, there are only two running coupling constants:
\begin{equation}
  w_h(\psi)=\sum_{i=0}^1\ell_i^{(h)}O_i(\psi)
\end{equation}
and
\begin{equation}
  O_0(\psi):=\frac12\sum_{\sigma,\sigma'\in\{\uparrow,\downarrow\}}\sum_{j=1,2,3}\psi_\sigma^+S^{(j)}_{\sigma,\sigma'}\psi_{\sigma'}^-
  \otimes S^{(j)}
  ,\quad
  O_1(\psi):=\frac12\left(\sum_{\sigma,\sigma'\in\{\uparrow,\downarrow\}}\sum_{j=1,2,3}\psi_\sigma^+S^{(j)}_{\sigma,\sigma'}\psi_{\sigma'}^-\right)^2
  \otimes\mathds 1
\end{equation}
(the $O$ are a product of Grassmann variables and a matrix acting on the impurity).
\bigskip

\indent
We compute the beta function in the same way as for graphene, see\-~(\ref{betadef}):
\begin{equation}
  \begin{largearray}
    C^{(h)}\left(1+
      \sum_{i=0}^1\ell_i^{(h-1)}O_i(\psi^{[\leqslant h-1]}(\Delta))
    \right)
    =\\\hfill=
    \prod_{\eta=\pm}
    \int P(d\psi^{[h]}(\Delta_\eta)
    \left(\sum_{i=0}^1\ell_i^{(h)}O_i\left(\psi^{[h]}(\Delta)+2^{-\gamma}\psi^{[\leqslant h-1]}(\Delta_\eta)\right)\right)^2
  \end{largearray}
\end{equation}
with $\gamma\equiv\frac12$.
In this expression, the square plays the same role as the prefactor $2^{d+1}$ in\-~(\ref{betadef}).
The power counting is determined by the sign of
\begin{equation}
  1+2l\gamma\equiv1+l
\end{equation}
so quadratic terms ($\ell_0$) are {\it marginal} and quartic terms ($\ell_1$) are {\it irrelevant}.
\bigskip

\begin{figure}
\hfil\includegraphics[width=12cm]{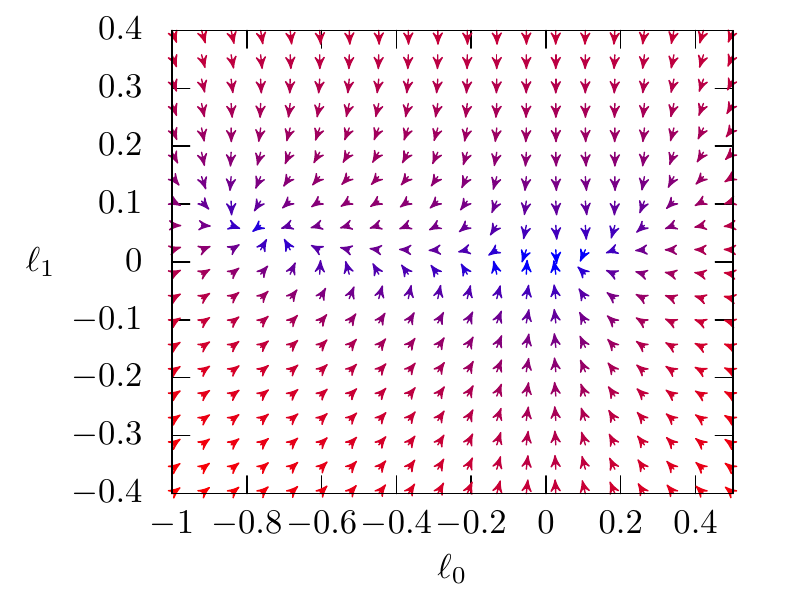}
\caption{
  The directional vector field of the beta function for the hierarchical Kondo model.
  (Each arrow shows the direction of the vector field, the color corresponds to the logarithm of the amplitude, with red being larger and blue smaller.)
  The stable and unstable equilibrium points are clearly visible.
}
\label{fig:sd_vector_field}
\end{figure}

\indent
The beta function can be computed exactly for the hierarchical model:
\begin{equation}
  \begin{array}{r@{\ }>{\displaystyle}l}
    C^{(h)}=&1+\frac32(\ell_0^{(h)})^2+9(\ell_1^{(h)})^2\\[0.3cm]
    \ell_0^{(h-1)}=&\frac1{C^{(h)}}\Big(\ell_0^{(h)}+3 \ell_0^{(h)}\ell_1 -(\ell_0^{(h)})^2\Big)\\[0.3cm]
    \ell_1^{(h-1)}=&\frac1{C^{(h)}}\Big( \frac12\ell_1^{(h)} +\frac18(\ell_0^{(h)})^2\Big)
    .
  \end{array}
\end{equation}
One can show that there are two equilibrium points: $\ell_0=\ell_1=0$, as well as
\begin{equation}
\ell^*_0=-x_0\frac{1+5x_0}{1-4x_0},\quad \ell^*_1=\frac{x_0}3
\label{kondoeqfixedpt}\end{equation}
where $x_0\approx0.15878626704216...$ is the real root of $4-19x-22x^2-107x^3=0$.
The trivial equilibrium at $(0,0)$ is stable if and only if $\ell_0>0$ (that is, if it is approached from the right).
Otherwise, the flow goes to the non-trivial equilibrium point, see figure\-~\ref{fig:sd_vector_field}
\bigskip

\bigskip

\indent
Thus, for the hierarchical Kondo model, the flow can go to a {\it non-trivial equilibrium point}, at which the interaction is preponderant.
Physically, this translates to the behavior of the interacting system being qualitatively different from the non-interacting one (in this case, this manifests itself by the magnetic susceptibility of the impurity being finite at $0$-temperature, which means that the impurity resists an external magnetic field even in the ground state, which is not the case for the non-interacting system).

\vfill
\eject
\appendix

\section{Free Fermions}\label{app:free_fermions}
Let
\begin{equation}
  \mathcal H_0=\sum_{i,j=1}^N \mu_{i,j}a^\dagger_ia_j
\end{equation}
where $(a_i)_i$ is a family of Fermionic annihilation operators and $\mu_{i,j}=\mu_{j,i}^*$.
Given an operator $A$ on Fock space and $\beta>0$, let
\begin{equation}
  \left<A\right>:=\frac{\mathrm{Tr}(e^{-\beta\mathcal H_0}A)}{\mathrm{Tr}(e^{-\beta\mathcal H_0})}
\end{equation}
and, for $t\in[0,\beta)$, let
\begin{equation}
  a_i^-(t):=e^{t\mathcal H_0}a_ie^{-t\mathcal H_0}
  ,\quad
  a_i^+(t):=e^{t\mathcal H_0}a_i^\dagger e^{-t\mathcal H_0}
  .
\end{equation}
Furthermore, we define the Fermionic time ordering operator as a linear operator on polynomials of $a_i(t)$ such that, for any $n\in\{1,\cdots,N\}$, $j_1,\cdots,j_n\in\{1,\cdots,N\}$, $\omega_1,\cdots,\omega_n\in\{-,+\}$, $t_1,\cdots,t_n\in[0,\beta)$,
\begin{equation}
  \mathbf T\left(\prod_{i=1}^na^{\omega_i}_{j_i}(t_i)\right)
  =
  (-1)^\tau\prod_{i=1}^na^{\omega_{\tau(i)}}_{j_{\tau(i)}}(t_{\tau(i)})
\end{equation}
where $\tau$ is the permutation of $\{1,\cdots,n\}$ that orders the $t_i$'s in reverse order, and in the case of equality of $t$'s, $\tau$ places $\omega=-$ to the left of $\omega=+$ and orders the $j$'s.
Formally, $\tau$ is such that $t_{\tau(i)}\geqslant t_{\tau(i+1)}$ and if $t_{\tau(i)}=t_{\tau(i+1)}$, then either $\omega_{\tau(i)}=-$ and $\omega_{\tau(i+1)}=+$ or $j_{\tau(i)}<j_{\tau(i+1)}$.
\bigskip

\indent
Let us first prove a technical lemma.
\bigskip
\theo{Lemma}\label{lemma:fock_technical}
  For $t\in\mathbb R$, and $\lambda_1,\cdots,\lambda_N\in\mathbb R$,
  \begin{equation}
    e^{-t\sum_{i=1}^N\lambda_ia_i^\dagger a_i}
    =
    \prod_{i=1}^N(1+(e^{-t\lambda_i}-1)a_i^\dagger a_i)
    \label{fock1}
  \end{equation}
  and
  \nopagebreakaftereq
  \begin{equation}
    e^{t\sum_{i=1}^N\lambda_ia_i^\dagger a_i}a_j^\dagger e^{-t\sum_{i=1}^N\lambda_ia_i^\dagger a_i}
    =e^{t\lambda_j}a_j^\dagger
    ,\quad
    e^{t\sum_{i=1}^N\lambda_ia_i^\dagger a_i}a_j e^{-t\sum_{i=1}^N\lambda_ia_i^\dagger a_i}
    =e^{-t\lambda_j}a_j
    .
    \label{fock4}
  \end{equation}
\endtheo
\restorepagebreakaftereq
\bigskip

\indent\underline{Proof}:
  To prove\-~(\ref{fock1}), we write $e^{-t\sum_i\lambda_ia_i^\dagger a_i}=\prod_ie^{-t\lambda_ia_i^\dagger a_i}$ and expand the exponential, using the fact that $(a_i^\dagger a_i)^n=a_i^\dagger a_i$ for any $n\geqslant 1$.
  By\-~(\ref{fock1}),
  \begin{equation}
    e^{-t\sum_{i=1}^N\lambda_ia_i^\dagger a_i}a_j^\dagger
    =
    \left(\prod_{i=1}^n(1+(e^{-t\lambda_i}-1)a_i^\dagger a_i)\right)a_j^\dagger
  \end{equation}
  and, commuting $a_j^\dagger$ through using $\{a_i,a_j^\dagger\}=\delta_{i,j}$ and $(a_j^\dagger)^2=0$, we find
  \begin{equation}
    e^{-t\sum_{i=1}^N\lambda_ia_i^\dagger a_i}a_j^\dagger
    =
    e^{-t\lambda_j}a_j^\dagger\prod_{i\neq j}
    (1+(e^{-t\lambda_i}-1)a_i^\dagger a_i)
  \end{equation}
  and since $(a_j^\dagger)^2=0$,
  \begin{equation}
    e^{-t\sum_{i=1}^N\lambda_ia_i^\dagger a_i}a_j^\dagger
    =
    e^{-t\lambda_j}a_j^\dagger\prod_{i}
    (1+(e^{-t\lambda_i}-1)a_i^\dagger a_i)
    =
    e^{-t\lambda_j}a_j^\dagger
    e^{-t\sum_{i=1}^N\lambda_ia_i^\dagger a_i}
    .
    \label{fock2}
  \end{equation}
  Taking the $\dagger$ of\-~(\ref{fock2}), we find
  \begin{equation}
    a_je^{-t\sum_{i=1}^N\lambda_ia_i^\dagger a_i}
    =
    e^{-t\lambda_j}
    e^{-t\sum_{i=1}^N\lambda_ia_i^\dagger a_i}
    a_j
    .
    \label{fock3}
  \end{equation}
  Combining\-~(\ref{fock2}) and\-~(\ref{fock3}), we find\-~(\ref{fock4}).
\qed
\bigskip

\subsection{Two-point correlation function}
\indent
We now compute the two-point correlation function of $\left<\cdot\right>$.
\bigskip

\theo{Lemma}\label{lemma:schwinger}
  For $i,j\in\{1,\cdots,N\}$ and $t,t'\in[0,\beta)$, if $t\neq t'$, then
  \begin{equation}
    s_{i,j}(t-t'):=\left<\mathbf T(a_i^-(t)a_j^+(t'))\right>
    =
    \frac1\beta\sum_{k_0\in\frac{2\pi}\beta(\mathbb Z+\frac12)}
    e^{-ik_0(t-t')}(-ik_0\mathds 1+\mu)^{-1}_{i,j}
    \label{schwinger}
  \end{equation}
  and if $t=t'$,
  \nopagebreakaftereq
  \begin{equation}
    s_{i,j}(0):=\left<\mathbf T(a_i^-(t)a_j^+(t))\right>
    =
    \frac12\delta_{i,j}+
    \frac1\beta\sum_{k_0\in\frac{2\pi}\beta(\mathbb Z+\frac12)}
    (-ik_0\mathds 1+\mu)^{-1}_{i,j}
    .
    \label{schwinger}
  \end{equation}
\endtheo
\restorepagebreakaftereq
\bigskip

\indent\underline{Proof}:
  We diagonalize $\mu$ by a unitary transform $U$, and define
  \begin{equation}
    b_i:=\sum_{j=1}^NU_{j,i}^*a_j
    ,\quad
    b_i^\dagger:=\sum_{j=1}^NU_{j,i}a_j^\dagger
  \end{equation}
  in terms of which
  \begin{equation}
    e^{-t\mathcal H_0}
    =
    e^{-t\sum_{i=1}^N\lambda_ib_i^\dagger b_i}
    =
    \prod_{i=1}^N(1+(e^{-t\lambda_i}-1)b_i^\dagger b_i)
  \end{equation}
  where $\lambda_i$ are the eigenvalues of $\mu$.
  Note that
  \begin{equation}
    \{b_i,b_j^\dagger\}
    =
    \sum_{k,l=1}^NU_{k,i}^*U_{l,j}\{a_k,a_l^\dagger\}
    =
    \sum_{k=1}^NU_{k,i}^*U_{k,j}
    =\delta_{i,j}
    .
  \end{equation}
  Thus
  \begin{equation}
    \mathrm{Tr}(e^{-\beta\mathcal H_0})
    =\prod_{i=1}^N(1+e^{-\beta\lambda_i})
    .
  \end{equation}
  Let
  \begin{equation}
    b_i^{-}(s):=e^{t\mathcal H_0}b_ie^{-t\mathcal H_0}
    =\sum_{j=1}^NU_{j,i}^*a_j^{-}(s)
    ,\quad
    b_i^{+}(s):=
    e^{t\mathcal H_0}b_i^\dagger e^{-t\mathcal H_0}
    =
    \sum_{j=1}^NU_{j,i}a_j^{+}(s)
    .
  \end{equation}
  By\-~(\ref{fock4}),
  \begin{equation}
    b_i^-(s)=e^{-s\lambda_i}b_i
    ,\quad
    b_i^+(s)=e^{s\lambda_i}b_i^\dagger
    .
  \end{equation}
  \bigskip

  \point
  Now, if $t\geqslant t'$,
  \begin{equation}
    \mathbf T(a_i^-(t)a_j^+(t'))
    =a_i^-(t)a_j^+(t')
    =\sum_{k,l=1}^N
    U_{i,k}U_{j,l}^*
    e^{-t\lambda_k+t'\lambda_l}b_kb_l^\dagger
    .
  \end{equation}
  In order for the trace to be non-zero, we must take $k=l$, so
  \begin{equation}
    \mathrm{Tr}(e^{-\beta\mathcal H_0}\mathbf T(a_i^-(t)a_j^+(t')))
    =\sum_{k=1}^N
    U_{i,k}U_{j,k}^*
    e^{-(t-t')\lambda_k}
    \mathrm{Tr}(e^{-\beta\mathcal H_0}b_kb_k^\dagger)
  \end{equation}
  and since $b_kb_k^\dagger=1-b_k^\dagger b_k$,
  \begin{equation}
    \mathrm{Tr}(e^{-\beta\mathcal H_0}\mathbf T(a_i^-(t)a_j^+(t')))
    =\sum_{k=1}^N
    U_{i,k}U_{j,k}^*
    e^{-(t-t')\lambda_k}
    \prod_{l\neq k}(1+e^{-\beta\lambda_l})
    .
  \end{equation}
  Therefore,
  \begin{equation}
    s_{i,j}(t-t')
    =
    \sum_{k=1}^N
    U_{i,k}U_{j,k}^*
    \frac{e^{-(t-t')\lambda_k}}{1+e^{-\beta\lambda_k}}
    .
    \label{S1}
  \end{equation}
  \bigskip

  \point If $t<t'$, then
  \begin{equation}
    \mathbf T(a_i^-(t)a_j^+(t'))
    =-a_j^+(t')a_i^-(t)
    =-e^{t'\mathcal H_0}a_j^\dagger e^{-(t'-t)\mathcal H_0}a_i e^{-t\mathcal H_0}
  \end{equation}
  so, using the cyclicity of the trace,
  \begin{equation}
    \mathrm{Tr}(e^{-\beta\mathcal H_0}\mathbf T(a_i^-(t)a_j^+(t')))
    =-\mathrm{Tr}(e^{t\mathcal H_0}a_i e^{-(t+\beta-t')\mathcal H_0}a_j^\dagger e^{-t'\mathcal H_0})
  \end{equation}
  which is identical to the trace computed above, but with $t$ replaced by $t+\beta$, so
  \begin{equation}
    s_{i,j}(t-t')
    =
    -\sum_{k=1}^NU_{i,k}U_{j,k}^*\frac{e^{-(t-t'+\beta)\lambda_k}}{1+e^{-\beta\lambda_k}}
    .
    \label{S2}
  \end{equation}
  \bigskip

  \point
  Let us compute the Fourier transform of $s_{i,j}(t-t')$ with respect to $t-t'\in(-\beta,\beta)$: for $k_0\in\frac\pi\beta\mathbb Z$, by\-~(\ref{S1}) and\-~(\ref{S2}),
  \begin{equation}
    \frac12\int_{-\beta}^\beta d\tau\ e^{ik_0\tau}s_{i,j}(t-t')
    =
    \sum_{k=1}^NU_{i,k}U_{j,k}^*
    \frac1{1+e^{-\beta\lambda_k}}
    \left(
      \int_0^\beta d\tau\ e^{ik_0\tau}e^{-\tau\lambda_k}
      -
      \int_{-\beta}^0 d\tau\ e^{ik_0\tau}e^{-(\tau+\beta)\lambda_k}
    \right)
  \end{equation}
  and so
  \begin{equation}
    \frac12\int_{-\beta}^\beta d\tau\ e^{ik_0\tau}s_{i,j}(t-t')
    =
    \sum_{k=1}^N
    \frac{U_{i,k}U_{j,k}^*}{1+e^{-\beta\lambda_k}}
    \left(
      \frac{e^{(ik_0-\lambda_k)\beta}-1}{ik_0-\lambda_k}
      -
      \frac{e^{-\beta\lambda_k}-e^{-ik_0\beta}}{ik_0-\lambda_k}
    \right)
    .
  \end{equation}
  Thus, if $k_0\in\frac\pi\beta2\mathbb Z$, then
  \begin{equation}
    \frac12\int_{-\beta}^\beta d\tau\ e^{ik_0\tau}s_{i,j}(t-t')
    =
    0
  \end{equation}
  and if $k_0\in\frac{2\pi}\beta(\mathbb Z+\frac12)$, then
  \begin{equation}
    \frac12\int_{-\beta}^\beta d\tau\ e^{ik_0\tau}s_{i,j}(t-t')
    =
    \sum_{k=1}^NU_{i,k}U_{j,k}^*
    \frac1{-ik_0+\lambda_k}
    .
  \end{equation}
  Thus, if the Fourier transform can be inverted, then we have
  \begin{equation}
    s_{i,j}(t-t')
    =
    \frac1\beta\sum_{k_0\in\frac{2\pi}\beta(\mathbb Z+\frac12)}
    e^{-ik_0(t-t')}(-ik_0\mathds 1+\mu)^{-1}_{i,j}
  \end{equation}
  and this is the case where $s_{i,j}(t-t')$ is continuous.
  \bigskip

  \point
  Obviously, $s_{i,j}$ is continuous at $t-t'\neq0$.
  At $t-t'=0$, we have
  \begin{equation}
    \lim_{t-t'\to 0_+}s_{i,j}(t-t')=\lim_{t-t'\to 0_-}s_{i,j}(t-t')+\delta_{i,j}
  \end{equation}
  so $s_{i,j}$ is continuous if $i\neq j$.
  Now, if $i=j$ and $t=t'$, then by\-~(\ref{S1}),
  \begin{equation}
    s_{i,j}(0)=\sum_{k=1}^NU_{i,k}U_{i,k}^*\frac1{1+e^{-\beta\lambda_k}}
    .
  \end{equation}
  Now,
  \begin{equation}
    s_{i,j}(t-t')-\frac12\mathrm{sgn}(t-t')
  \end{equation}
  is continuous at $t-t'=0$, and its Fourier transform is, by a similar computation to that above,
  \begin{equation}
    \frac12\int_{-\beta}^\beta d\tau\ \left(s_{i,j}(\tau)-\frac12\mathrm{sgn}(\tau)\right)
    =
    \sum_{k=1}^NU_{i,k}U_{j,k}^*
    \left(\frac1{-ik_0+\lambda_k}+\frac1{ik_0}\right)
  \end{equation}
  which is absolutely summable in $k_0$.
  Thus
  \begin{equation}
    \lim_{\tau\to0_+}\frac1\beta\sum_{k_0\in\frac{2\pi}\beta(\mathbb Z+\frac12)}
    e^{-ik_0\tau}
    \left((-ik_0\mathds 1+\mu)^{-1}+\frac1{ik_0}\mathds 1\right)
    =
    \frac1\beta\sum_{k_0\in\frac{2\pi}\beta(\mathbb Z+\frac12)}
    \left((-ik_0\mathds 1+\mu)^{-1}+\frac1{ik_0}\mathds 1\right)
  \end{equation}
  and so
  \begin{equation}
    \lim_{\tau\to0_+}\left(s_{i,j}(\tau)-\frac12\mathrm{sgn}(\tau)\right)
    =
    \frac1\beta\sum_{k_0\in\frac{2\pi}\beta(\mathbb Z+\frac12)}
    (-ik_0\mathds 1+\mu)^{-1}
  \end{equation}
  (the sum is to be understood as a principal part: $\sum_{k_0}(-ik_0\mathds 1+\mu)^{-1}\equiv \sum_{k_0}\mu(k_0^2\mathds 1+\mu^2)^{-1}$) and thus
  \begin{equation}
    s_{i,j}(0)=\frac12
    +\frac1\beta\sum_{k_0\in\frac{2\pi}\beta(\mathbb Z+\frac12)}
    (-ik_0\mathds 1+\mu)^{-1}
    .
  \end{equation}
\qed

\subsection{Wick rule}
\indent
This system of {\it free Fermions} satisfies the Wick rule, stated below.
\bigskip

\theoname{Lemma}{Wick rule}\label{lemma:wick}
  For any $n\in\{1,\cdots,N\}$, $j_1,\cdots,j_n\in\{1,\cdots,N\}$, $\bar j_1,\cdots,\bar j_n\in\{1,\cdots,N\}$, $t_1,\cdots,t_n\in[0,\beta)$, $\bar t_1,\cdots,\bar t_n\in[0,\beta)$,
  \nopagebreakaftereq
  \begin{equation}
    \left<\mathbf T\left(\prod_{i=1}^{n}a_{j_i}^-(t_i)a_{\bar j_i}^+(\bar t_i)\right)\right>
    =
    \sum_{\tau\in\mathcal S_n}(-1)^\tau
    \prod_{i=1}^{n}
    \left<\mathbf T\left(a_i^-(t_i)a_{\bar j_{\tau(i)}}^+(\bar t_{\tau(i)})\right)\right>
    .
    \label{wick_app}
  \end{equation}
\endtheo
\restorepagebreakaftereq
\bigskip

\indent\underline{Proof}:
  First, note that
  \begin{equation}
    \left<\mathbf T\left(\prod_{i=1}^{n}a_{j_i}^-(t_i)a_{\bar j_i}^+(\bar t_i)\right)\right>
    =
    \left<\mathbf T\left(\left(\prod_{i=1}^{n}a_{j_i}^-(t_i)\right)\left(\prod_{i=n}^1a_{\bar j_i}^+(\bar t_i)\right)\right)\right>
    .
    \label{fock_split_wick}
  \end{equation}
  Without loss of generality, we may assume that $t_1\geqslant\cdots\geqslant t_n$ and $\bar t_1\geqslant\cdots\geqslant\bar t_n$ (indeed, (\ref{wick_app}) is antisymmetric under exchanges of $a^{-}$'s and exchanges of $a^+$'s).
  \bigskip

  \point
  We diagonalize $\mu$ by a unitary transform $U$, and define
  \begin{equation}
    b_i:=\sum_{j=1}^NU_{j,i}^*a_j
    ,\quad
    b_i^\dagger:=\sum_{j=1}^NU_{j,i}a_j^\dagger
  \end{equation}
  in terms of which
  \begin{equation}
    e^{-t\mathcal H_0}
    =
    e^{-t\sum_{i=1}^N\lambda_ib_i^\dagger b_i}
    =
    \prod_{i=1}^N(1+(e^{-t\lambda_i}-1)b_i^\dagger b_i)
  \end{equation}
  where $\lambda_i$ are the eigenvalues of $\mu$.
  Note that
  \begin{equation}
    \{b_i,b_j^\dagger\}
    =
    \sum_{k,l=1}^NU_{k,i}^*U_{l,j}\{a_k,a_l^\dagger\}
    =
    \sum_{k=1}^NU_{k,i}^*U_{k,j}
    =\delta_{i,j}
  \end{equation}
  Let
  \begin{equation}
    b_i^{-}(s):=e^{t\mathcal H_0}b_ie^{-t\mathcal H_0}
    =\sum_{j=1}^NU_{j,i}^*a_j^{-}(s)
    ,\quad
    b_i^{+}(s):=
    e^{t\mathcal H_0}b_i^\dagger e^{-t\mathcal H_0}
    =
    \sum_{j=1}^NU_{j,i}a_j^{+}(s)
    .
  \end{equation}
  By\-~(\ref{fock4}),
  \begin{equation}
    b_i^-(s)=e^{-s\lambda_i}b_i
    ,\quad
    b_i^+(s)=e^{s\lambda_i}b_i^\dagger
    .
  \end{equation}
  We have
  \begin{equation}
    \begin{largearray}
      \left<\mathbf T\left(\left(\prod_{i=1}^{n}a_{j_i}^-(t_i)\right)\left(\prod_{i=n}^1a_{\bar j_i}^+(\bar t_i)\right)\right)\right>
      =\\\hfill=
      \sum_{k_1,\cdots,k_n=1}^N\sum_{l_1,\cdots,l_n=1}^N
      \left(\prod_{i=1}^nU_{j_i,k_i}U_{\bar j_i,l_i}^*\right)
      \left<\mathbf T\left(\left(\prod_{i=1}^{n}b_{k_i}^-(t_i)\right)\left(\prod_{i=n}^1b_{l_i}^+(\bar t_i)\right)\right)\right>
      .
    \end{largearray}
    \label{fock_diagonalization}
  \end{equation}
  \bigskip

  \point
  Let $\sigma\in\{1,-1\}$ and $\gamma_1,\cdots,\gamma_{2n}\in\{1,\cdots,N\}$, $\omega_1,\cdots,\omega_{2n}\in\{+,-\}$, $\beta\geqslant s_1\geqslant\cdots\geqslant s_{2n}\geqslant0$ be such that
  \begin{equation}
    \mathbf T\left(\left(\prod_{i=1}^{n}b_{k_i}^-(t_i)\right)\left(\prod_{i=n}^1b_{l_i}^+(\bar t_j)\right)\right)
    =
    \sigma\prod_{i=1}^{2n}b_{\gamma_i}^{\omega_i}(s_i)
  \end{equation}
  and, by\-~(\ref{fock4}),
  \begin{equation}
    \mathbf T\left(\left(\prod_{i=1}^{n}b_{k_i}^-(t_i)\right)\left(\prod_{i=n}^1b_{l_i}^+(\bar t_j)\right)\right)
    =
    \left(\prod_{i=1}^n e^{-\lambda_{k_i}t_i+\lambda_{l_i}\bar t_i}\right)
    \sigma\prod_{i=1}^{2n}b_{\gamma_i}^{\omega_i}
    \label{fock_timeorder}
  \end{equation}
  where $b_\gamma^+\equiv b_\gamma^\dagger$ and $b_\gamma^-\equiv b_\gamma$.
  We now commute the $b_\gamma$ back into the order $\prod_{i=1}^nb_{k_i}\prod_{i=n}^1b_{l_i}^\dagger$, in such a way that the $\sigma$ sign is canceled.
  This is not entirely straightforward: whenever a $b^\dagger_{l_i}$ is commuted with a $b_{k_j}$, it either moves through or yields $\delta_{l_i,k_j}$.
  To keep track of these terms, we introduce the following notation: given a product of $b$'s, we define
  \begin{equation}
    \left|b^\dagger_{l_n}\cdots b^\dagger_{l_1}\right\}
    :=
    \left(\prod_{i=1}^nb_{k_i}\right)\left(\prod_{i=n}^1b^\dagger_{l_i}\right)
  \end{equation}
  and make the symbol $\left|\cdot\right\}$ multilinear: for instance,
  \begin{equation}
    \left|(b^\dagger_{l_n}-1)\cdots b^\dagger_{l_1}\right\}
    \equiv
    \left(\prod_{i=1}^nb_{k_i}\right)\left(\prod_{i=n}^1b^\dagger_{l_i}\right)
    -
    \left(\prod_{i=1}^{n-1}b_{k_i}\right)\left(\prod_{i=n-1}^1b^\dagger_{l_i}\right)
    .
  \end{equation}
  We move the $b^\dagger$ to the right of the $b$ and put them back in the order $b_{l_n}^\dagger\cdots b_{l_1}^\dagger$.
  Note that, since $\bar t_1\geqslant\cdots\bar t_n$ and $t_1\geqslant\cdots\geqslant t_n$, the $b_i$ are already ordered as in $\prod_{i=1}^nb_i$, but the $b_i^\dagger$ are ordered in the opposite order as in $\prod_{i=n}^1b_i^\dagger$.
  As we will now discuss, reordering the $b$'s in this way yields
  \begin{equation}
    \sigma\prod_{i=1}^{2n}b_{\gamma_i}^{\omega_i}
    =
    \left|\prod_{i=n}^1\left(b_{l_i}^\dagger-{\textstyle\sum_{j=1}^n}(-1)^{j+i}\delta_{l_i,k_j}\mathds 1_{\bar t_i> t_j}\right)\right\}
    \label{fock_reorder}
  \end{equation}
  where $\mathds 1_{\bar t_i> t_j}\in\{0,1\}$ is equal to 1 if and only if $\bar t_i> t_j$.
  When moving the $b^\dagger$'s, the operators either anti-commute, or, when passing $b_i^\dagger$ over $b_i$, the two operators may destroy each other (the $\delta_{i,j}$ term in $\{a_i,a_j^\dagger\}=\delta_{i,j}$).
  When the operators destroy each other, they become numbers, and commute with all other operators.
  After this, they no longer produce the signs necessary to cancel out $\sigma$.
  The sign in $\sigma$ due to those terms is $-1$ to the power of the number of positions between the location where the operators were destroyed and the location where they end up.
  Because of the ordering of the $t$'s and $\bar t$'s, if $b^\dagger_{l_i}$ and $b_{k_j}$ destroyed each other, then there are $n-j$ $b$'s left to go over and $n-i$ $b^\dagger$'s.
  Thus, the contribution of $\sigma$ for this term is $(-1)^{n-j+n-i}=(-1)^{i+j}$.
  This justifies\-~(\ref{fock_reorder}).
  \bigskip

  \point
  We show that
  \begin{equation}
    \left<\left|\prod_{i=n}^1b_{l_i}^\dagger\right\}\right>
    \equiv
    \left<\left(\prod_{i=1}^nb_{k_i}\right)\left(\prod_{i=n}^1b_{l_i}^\dagger\right)\right>
    =
    \sum_{\tau\in\mathcal S_n}(-1)^\tau\prod_{i=1}^n\left<b_{k_i}b_{l_{\tau(i)}}^\dagger\right>
    .
    \label{wick_notime}
  \end{equation}
  First of all, $\prod_{i=n}^1b_{l_i}^\dagger$ is non-zero only if all $l_i$ are different from each other, and so is the right side of\-~(\ref{wick_notime}), so we can assume that all $l_i$ are different from each other.
  Next, since $\mathcal H_0=\sum_{\lambda_i}b_i^\dagger b_i$, the $b_{l_i}^\dagger$ and $b_{k_i}$ must be paired up.
  Using the antisymmetry of both sides of\-~(\ref{wick_notime}), we can assume without loss of generality that $k_i=l_i$.
  In that case,
  \begin{equation}
    \mathrm{Tr}\left(e^{-\beta\mathcal H_0}\left(\prod_{i=1}^nb_{k_i}\right)\left(\prod_{i=n}^1b_{l_i}^\dagger\right)\right)
    =
    \mathrm{Tr}\left(e^{-\beta\mathcal H_0}\left(\prod_{i=1}^nb_{k_i}\right)\left(\prod_{i=n}^1b_{k_i}^\dagger\right)\right)
    =
    \mathrm{Tr}\left(e^{-\beta\mathcal H_0}\prod_{i=1}^n b_{k_i}b_{k_i}^\dagger\right)
  \end{equation}
  and so
  \begin{equation}
    \left<e^{-\beta\mathcal H_0}\left(\prod_{i=1}^nb_{k_i}\right)\left(\prod_{i=n}^1b_{l_i}^\dagger\right)\right>
    =
    \prod_{i=1}^n\left<b_{k_i}b_{k_i}^\dagger\right>
  \end{equation}
  which is equal to the right side of\-~(\ref{wick_notime}).
  \bigskip

  \point
  By\-~(\ref{wick_notime}),
  \begin{equation}
    \left<\left|\prod_{i=n}^1\left(b_{l_i}^\dagger-{\textstyle\sum_{j=1}^n}(-1)^{i+j}\delta_{l_i,k_j}\mathds 1_{\bar t_i> t_j}\right)\right\}\right>
    =
    \sum_{\tau\in\mathcal S_n}(-1)^\tau\prod_{i=1}^n\left(\left<b_{k_i}b_{l_{\tau(i)}}^\dagger\right>-\delta_{k_i,l_{\tau(i)}}\mathds 1_{\bar t_{\tau(i)}> t_i}\right)
    .
    \label{fock_det}
  \end{equation}
  Now,
  \begin{equation}
    \left<b_{k_i}b_{l_{\tau(i)}}^\dagger\right>
    =
    \delta_{k_i,l_{\tau(i)}}
    \frac1{1+e^{-\beta\lambda_{k_i}}}
    \label{2pt_notime_+}
  \end{equation}
  and
  \begin{equation}
    \left<b_{k_i}b_{l_{\tau(i)}}^\dagger\right>
    -\delta_{k_i,l_{\tau(i)}}
    =
    -\delta_{k_i,l_{\tau(i)}}
    \frac{e^{-\beta\lambda_{k_i}}}{1+e^{-\beta\lambda_{k_i}}}
    .
    \label{2pt_notime_-}
  \end{equation}
  Thus, inserting\-~(\ref{2pt_notime_+}), (\ref{2pt_notime_-}) into\-~(\ref{fock_det}), (\ref{fock_reorder}), (\ref{fock_timeorder}), (\ref{fock_diagonalization}) and\-~(\ref{fock_split_wick}), we find
  \begin{equation}
    \begin{largearray}
      \left<\mathbf T\left(\prod_{i=1}^{n}a_{j_i}^-(t_i)a_{\bar j_i}^+(\bar t_i)\right)\right>
      =\\\hfill=
      \sum_{k_1,\cdots,k_n=1}^N\sum_{l_1,\cdots,l_n=1}^N
      \left(\prod_{i=1}^nU_{j_i,k_i}U_{\bar j_i,l_i}^*\right)
      \sum_{\tau\in\mathcal S_n}(-1)^\tau\prod_{i=1}^n
      \delta_{k_i,l_{\tau(i)}}e^{-\lambda_{k_i}(t_i-\bar t_{\tau(i)})}
      \frac{-\mathds 1_{\bar t_{\tau(i)}> t_i}e^{-\beta\lambda_{k_i}}+\mathds 1_{\bar t_{\tau(i)}<t_i}}{1-e^{-\beta\lambda_{k_i}}}
    \end{largearray}
  \end{equation}
  and, relabeling the $l_i$,
  \begin{equation}
    \begin{largearray}
      \left<\mathbf T\left(\prod_{i=1}^{n}a_{j_i}^-(t_i)a_{\bar j_i}^+(\bar t_i)\right)\right>
      =\\\hfill=
      \sum_{\tau\in\mathcal S_n}(-1)^\tau
      \prod_{i=1}^n\left(
	\sum_{k=1}^N
	U_{j_i,k}U_{\bar j_{\tau(i)},k}^*
	e^{-\lambda_{k}(t_i-\bar t_{\tau(i)})}
	\frac{-\mathds 1_{\bar t_{\tau(i)}> t_i}e^{-\beta\lambda_{k}}+\mathds 1_{\bar t_{\tau(i)}<t_i}}{1-e^{-\beta\lambda_{k}}}
      \right)
    \end{largearray}
  \end{equation}
  which, by\-~(\ref{S1}) and\-~(\ref{S2}), is
  \begin{equation}
    \left<\mathbf T\left(\prod_{i=1}^{n}a_{j_i}^-(t_i)a_{\bar j_i}^+(\bar t_i)\right)\right>
    \sum_{\tau\in\mathcal S_n}(-1)^\tau
    \prod_{i=1}^n
    \left<\mathbf T(a_{j_i}^-(t_i)a_{\bar j_{\tau(i)}}^+(\bar t_{\tau(i)}))\right>
    .
  \end{equation}
\qed

\section{Properties of Gaussian Grassman integrals}\label{app:grassmann}
\indent
Consider a family $\psi_1^\pm,\cdots,\psi_N^\pm$ of Grassmann variables and the Gaussian Grassmann measure
\begin{equation}
  P_\mu(d\psi)=\det(\mu)\prod_{i=1}^Nd\psi_i^+d\psi_i^-
  e^{-\sum_{i,j=1}^N\mu_{i,j}\psi_i^+\mu^{-1}_{i,j}\psi_i^-}
\end{equation}
where $\mu$ is a Hermitian, invertible matrix.
\bigskip

\indent
We first prove that one can perform changes of variables in Grassmann integrals.

\theoname{Lemma}{Change of variables in Grassmann integrals}\label{lemma:change_vars_grassmann}
  Given a unitary $N\times N$ matrix $U$ and a polynomial $f$ in the Grassmann variables $\psi$, if
  \begin{equation}
    \varphi_i^-:=\sum_{j=1}^NU_{i,j}\psi_j^-
    ,\quad
    \varphi_i^+:=\sum_{j=1}^NU_{i,j}^*\psi_j^+
  \end{equation}
  then
  \nopagebreakaftereq
  \begin{equation}
    \int\prod_{i=1}^Nd\varphi_i^+d\varphi_i^-\ f(U^\dagger\varphi^-,(U^\dagger)^*\varphi^+)
    =
    \int \prod_{i=1}^Nd\psi_i^+d\psi_i^-\ f(\psi^-,\psi^+)
    .
  \end{equation}
\endtheo
\restorepagebreakaftereq
\bigskip

\indent\underline{Proof}:
  Without loss of generality, we can assume that the highest order term of $f$ is
  \begin{equation}
    \alpha\prod_{i=1}^N\psi_i^-\psi_i^+
    .
  \end{equation}
  We have
  \begin{equation}
    \int \prod_{i=1}^Nd\psi_i^+d\psi_i^-\ f(\psi^-,\psi^+)=\alpha
    .
  \end{equation}
  Furthermore,
  \begin{equation}
    \int\prod_{i=1}^Nd\varphi_i^+d\varphi_i^-\ f(U^\dagger\varphi^-,U\varphi^+)
    =
    \alpha\int\prod_{i=1}^Nd\varphi_i^+d\varphi_i^-\ 
    \prod_{i=1}^N\left(
      \sum_{j,k=1}^N
      U_{i,j}^*\varphi_j^-
      U_{i,k}\varphi_k^+
    \right)
  \end{equation}
  and, avoiding repetitions in the products, we find that
  \begin{equation}
    \prod_{i=1}^N\left(
      \sum_{j,k=1}^N
      U_{i,j}^*\varphi_j^-
      U_{i,k}\varphi_k^+
    \right)
    =
    \sum_{\tau,\tau'\in\mathcal S_N}
    \prod_{i=1}^NU^*_{i,\tau(i)}U_{i,\tau'(i)}\varphi_{\tau(i)}^-\varphi_{\tau'(i)}^+
  \end{equation}
  and so, reordering the $\varphi$,
  \begin{equation}
    \prod_{i=1}^N\left(
      \sum_{j,k=1}^N
      U_{i,j}^*\varphi_j^-
      U_{i,k}\varphi_k^+
    \right)
    =
    \sum_{\tau,\tau'\in\mathcal S_N}
    (-1)^\tau(-1)^{\tau'}
    \prod_{i=1}^NU^*_{i,\tau(i)},U_{i,\tau'(i)}\varphi_{i}^-\varphi_{i}^+
    =|\det(U)|^2
    \prod_{i=1}^N\varphi_i^-\varphi_i^+
  \end{equation}
  and, since $\det(U)=1$,
  \begin{equation}
    \int\prod_{i=1}^Nd\varphi_i^+d\varphi_i^-\ f(U^\dagger\varphi^-,U\varphi^+)
    =
    \alpha
    =
    \int \prod_{i=1}^Nd\psi_i^+d\psi_i^-\ f(\psi^-,\psi^+)
    .
  \end{equation}
\qed
\bigskip

\subsection{Gaussian Grassmann integrals}
\indent
Let us now turn to a set of identities for Gaussian Grassmann integrals.
\bigskip

\theo{Lemma}\label{lemma:grassmann_id}
  We have
  \begin{equation}
    \int P_\mu(d\psi)\ 1 = 1
  \end{equation}
  and
  \nopagebreakaftereq
  \begin{equation}
    \int P_\mu(d\psi)\ \psi_i^-\psi_j^+=\mu_{i,j}
  \end{equation}
  where $\mu^{-1}$ is the inverse matrix of $\mu$.
\endtheo
\restorepagebreakaftereq
\bigskip

\indent\underline{Proof}:
  This is a consequence of the change of variables lemma\-~\ref{lemma:change_vars_grassmann}.
  We diagonalize $\mu^{-1}$ by a unitary transform $U$, and change variables to $\varphi^-_i=\sum_{j}U_{j,i}^*\psi^-_j$ and $\varphi^+_i=\sum_j U_{j,i}\psi^+_j$:
  \begin{equation}
    \int P_\mu(d\psi)\ 1
    =
    \det(\mu)\int \prod_{i=1}^Nd\varphi_i^+d\varphi_i^-\ e^{-\sum_{i=1}^N\lambda_i\varphi_i^+\varphi_i^-}
  \end{equation}
  where $\lambda_i$ are the eigenvalues of $\mu$.
  Furthermore,
  \begin{equation}
    e^{-\sum_{i=1}^N\lambda_i\varphi_i^+\varphi_i^-}
    =
    \prod_{i=1}^Ne^{-\lambda_i\varphi_i^+\varphi_i^-}
    =
    \prod_{i=1}^N(1-\lambda_i\varphi_i^+\varphi_i^-)
    \label{exp_trivial}
  \end{equation}
  so
  \begin{equation}
    \int P_\mu(d\psi)\ 1
    =
    \det(\mu)\int \prod_{i=1}^Nd\varphi_i^+d\varphi_i^-\ 
    \prod_{i=1}^N(1+\lambda_i\varphi_i^-\varphi_i^+)
    =
    \det(\mu)\prod_{i=1}^N\lambda_i=1
    .
  \end{equation}
  \bigskip

  \indent
  Similarly,
  \begin{equation}
    \int P_\mu(d\psi)\ \psi_i^-\psi_j^+
    =
    \sum_{k,l=1}^N
    \det(\mu)\int \prod_{i=1}^Nd\varphi_i^+d\varphi_i^-\ e^{-\sum_{i=1}^N\lambda_i\varphi_i^+\varphi_i^-}
    U_{i,k}\varphi_k^-U_{j,l}^*\varphi_l^+
  \end{equation}
  in which we insert\-~(\ref{exp_trivial}) to find that $k$ must be equal to $l$ and
  \begin{equation}
    \int P_\mu(d\psi)\ 1
    =
    \det(\mu)
    \sum_{k=1}^N
    U_{k,j}^*
    U_{i,k}
    \lambda_k^{-1}
    \prod_{i=1}^N\lambda_i
    =
    \sum_{k=1}^N
    U_{j,k}^*
    U_{i,k}
    \lambda_k^{-1}
    =\mu_{i,j}
    .
  \end{equation}
\qed

\subsection{Wick rule for Gaussian Grassmann integrals}

\indent
Let us now turn to the Wick rule for Gaussian Grassmann integrals.
\bigskip

\theoname{Lemma}{Wick rule for Gaussian Grassmann integrals}\label{lemma:wick_grassmann}
  For $n\in\{1,\cdots,N\}$, $j_1,\cdots,j_n\in\{1,\cdots,N\}$ and $\bar j_1,\cdots,\bar j_n\in\{1,\cdots,N\}$,
  \nopagebreakaftereq
  \begin{equation}
    \int P_\mu(d\psi)\ 
    \prod_{i=1}^{n}\psi_{j_i}^-\psi_{\bar j_i}^+
    =
    \sum_{\tau\in\mathcal S_n}(-1)^\tau
    \prod_{i=1}^{n}
    \int P_\mu(d\psi)\ 
    \psi_{j_i}^-\psi_{\bar j_{\tau(i)}}^+
    .
  \end{equation}
\endtheo
\restorepagebreakaftereq
\bigskip

\indent\underline{Proof}:
  As before, we change variables using lemma\-~\ref{lemma:change_vars_grassmann}.
  We diagonalize $\mu^{-1}$ by a unitary transform $U$, and change variables to $\varphi^-_i=\sum_{j}U_{j,i}^*\psi^-_j$ and $\varphi^+_i=\sum_j U_{j,i}\psi^+_j$:
  \begin{equation}
    \begin{largearray}
      \int P_\mu(d\psi)\ 
      \prod_{i=1}^{n}\psi_{j_i}^-\psi_{\bar j_i}^+
      =\\\hfill=
      \det(\mu)
      \sum_{k_1,\cdots,k_n=1}^N
      \sum_{l_1,\cdots,l_n=1}^N
      \int \prod_{i=1}^Nd\varphi_i^+d\varphi_i^-\ 
      \left(\prod_{i=1}^n(1+\lambda_i\varphi_i^-\varphi_i^+)\right)
      \prod_{i=1}^{n}U_{j_i,k_i}\varphi_{k_i}^-U_{\bar j_i,l_i}^*\varphi_{l_i}^+
      .
    \end{largearray}
  \end{equation}
  Now, $\prod_i\varphi_{k_i}^-\varphi_{l_i}^+$ must be permutable into products of pairs with identical indices, otherwise the resulting integral would be 0.
  Summing over all such possibilities, we find
  \begin{equation}
    \begin{largearray}
      \int P_\mu(d\psi)\ 
      \prod_{i=1}^{n}\psi_{j_i}^-\psi_{\bar j_i}^+
      =\det(\mu)
      \cdot\\\hfill\cdot
      \sum_{k_1,\cdots,k_n=1}^N
      \sum_{\tau\in\mathcal S_n}(-1)^\tau
      \left(\prod_{i=1}^{n}U_{j_i,k_i}U_{\bar j_{\tau(i)},k_i}^*\right)
      \int \prod_{i=1}^Nd\varphi_i^+d\varphi_i^-\ 
      \left(\prod_{i=1}^n(1+\lambda_i\varphi_i^-\varphi_i^+)\right)
      \prod_{i=1}^n\varphi_{k_i}^-\varphi_{k_i}^+
    \end{largearray}
  \end{equation}
  and so
  \begin{equation}
    \int P_\mu(d\psi)\ 
    \prod_{i=1}^{n}\psi_{j_i}^-\psi_{\bar j_i}^+
    =
    \sum_{\tau\in\mathcal S_n}(-1)^\tau
    \prod_{i=1}^{n}\left(
      \sum_{k=1}^N
      U_{j_i,k}U_{\bar j_{\tau(i)},k}^*
      \lambda_{k}^{-1}
    \right)
    =
    \sum_{\tau\in\mathcal S_n}(-1)^\tau
    \prod_{i=1}^{n}\mu_{j_i,j_{\tau(i)}}
    .
  \end{equation}
\qed

\subsection{Addition property}
\indent
Finally, let us prove the addition property of Gaussiann Grassmann integrals.
\bigskip

\theoname{Lemma}{Addition property}\label{lemma:grassmann_addition}
  Consider two invertible matrices $\mu_1$ and $\mu_2$ and its two associated Gaussian Grassmann measure $P_{\mu_1}(d\psi_1)$ and $P_{\mu_2}(d\psi_2)$.
  For any polynomial $f$,
  \nopagebreakaftereq
  \begin{equation}
    \int P_{\mu_1+\mu_2}(d\psi)\ f(\psi)
    =
    \int P_{\mu_1}(d\psi_1)\int P_{\mu_2}(d\psi_2)\ f(\psi_1+\psi_2)
    .
  \end{equation}
\endtheo
\restorepagebreakaftereq
\bigskip

\indent\underline{Proof}:
  It is sufficient to prove the lemma when $f$ is a monomial of the form
  \begin{equation}
    f(\psi)=
    \prod_{i=1}^{n}
    \psi_{j_i}^-\psi_{\bar j_i}^+
    .
  \end{equation}
  First, note that $P_{\mu_1}(d\psi_1)P_{\mu_2}(d\psi_2)$ is proportional to a Gaussian Grassmann measure.
  Next, change variables to $\varphi_1=\frac1{\sqrt2}(\psi_1+\psi_2)$ and $\varphi_2=\frac1{\sqrt2}(\psi_1-\psi_2)$ in terms of which
  \begin{equation}
    \int P_{\mu_1}(d\psi_1)\int P_{\mu_2}(d\psi_2)\ f(\psi_1+\psi_2)
    =
    \int P_{\nu_1}(d\varphi_1)\int P_{\nu_2}(d\varphi_2)\ f(\varphi_1)
  \end{equation}
  where $\nu_1$ and $\nu_2$ can be computed from the change of variables, but this is not necessary here.
  Since $f$ is a monomial, it can be computed using the Wick rule\-~\ref{lemma:wick_grassmann}, and thus, changing variables back to $\psi$,
  \begin{equation}
    \begin{largearray}
      \int P_{\mu_1}(d\psi_1)\int P_{\mu_2}(d\psi_2)\ f(\psi)
      =\\\hfill=
      \sum_{\tau\in\mathcal S_n}(-1)^\tau
      \prod_{i=1}^{n}
      \int P_{\mu_1}(d\psi_1)\int P_{\mu_2}(d\psi_2)\ 
      (\psi_{1,j_i}^-+\psi_{2,j_i}^-)(\psi_{1,\bar j_{\tau(i)}}^++\psi_{2,\bar j_{\tau(i)}}^+)
      .
    \end{largearray}
  \end{equation}
  Now,
  \begin{equation}
    \begin{largearray}
      \int P_{\mu_1}(d\psi_1)\int P_{\mu_2}(d\psi_2)\ 
      (\psi_{1,j_i}^-+\psi_{2,j_i}^-)(\psi_{1,\bar j_{\tau(i)}}^++\psi_{2,\bar j_{\tau(i)}}^+)
      =\\\hfill=
      \int P_{\mu_1}(d\psi_1)\ 
      \psi_{1,j_i}^-\psi_{1,\bar j_{\tau(i)}}^+
      +
      \int P_{\mu_2}(d\psi_2)\ 
      \psi_{2,j_i}^-\psi_{2,\bar j_{\tau(i)}}^+
    \end{largearray}
  \end{equation}
  so, by lemma\-~\ref{lemma:grassmann_id},
  \begin{equation}
    \int P_{\mu_1}(d\psi_1)\int P_{\mu_2}(d\psi_2)\ 
    (\psi_{1,j_i}^-+\psi_{2,j_i}^-)(\psi_{1,\bar j_{\tau(i)}}^++\psi_{2,\bar j_{\tau(i)}}^+)
    =
    (\mu_1+\mu_2)_{j_i,\bar j_{\tau(i)}}
  \end{equation}
  and
  \begin{equation}
    \int P_{\mu_1+\mu_2}(d\psi)\ 
    \psi_{j_i}^-\psi_{\bar j_{\tau(i)}}^+
    =
    (\mu_1+\mu_2)_{j_i,\bar j_{\tau(i)}}
    .
  \end{equation}
  We conclude the proof of the lemma using the Wick rule for $\psi$:
  \begin{equation}
    \int P_{\mu_1+\mu_2}(d\psi)\ f(\psi)
    =
    \sum_{\tau\in\mathcal S_n}(-1)^\tau
    \prod_{i=1}^{n}
    \int P_{\mu_1+\mu_2}(d\psi)\ 
    \psi_{j_i}^-\psi_{\bar j_{\tau(i)}}^+
    .
  \end{equation}
\qed

\vfill
\eject

\end{document}